\documentclass[manuscript]{aastex}
\usepackage{epsf}
\usepackage{psfig}
\usepackage{graphicx}
\usepackage{lscape}
\usepackage{natbib}
\usepackage{enumerate}
\usepackage[flushleft]{threeparttable}
\usepackage{soul}
\usepackage{amsmath}

\shorttitle{}
\shortauthors{Rich et al.}

\begin{document}
\title{Multi-Epoch Direct Imaging and Time-Variable Scattered Light Morphology of the HD 163296 Protoplanetary Disk}
\author{Evan A. Rich\altaffilmark{1},
John P. Wisniewski\altaffilmark{1},
Thayne Currie\altaffilmark{3,5,46},
Misato Fukagawa\altaffilmark{6,28},
Carol A. Grady\altaffilmark{2,3,4},
Michael L. Sitko\altaffilmark{32,33},
Monika Pikhartova\altaffilmark{32},
Jun Hashimoto\altaffilmark{47},
Lyu Abe\altaffilmark{7},
Wolfgang Brandner\altaffilmark{8},
Timothy D. Brandt\altaffilmark{9},
Joseph C. Carson\altaffilmark{10},
Jeffrey Chilcote\altaffilmark{54},
Ruobing Dong\altaffilmark{55},
Markus Feldt\altaffilmark{8},
Miwa Goto\altaffilmark{11},
Tyler Groff\altaffilmark{50},
Olivier Guyon\altaffilmark{5,48,49},
Yutaka Hayano\altaffilmark{5},
Masahiko Hayashi\altaffilmark{6},
Saeko S. Hayashi\altaffilmark{5},
Thomas Henning\altaffilmark{8},
Klaus W. Hodapp\altaffilmark{12},
Miki Ishii\altaffilmark{6},
Masanori Iye\altaffilmark{6},
Markus Janson\altaffilmark{37},
Nemanja Jovanovic\altaffilmark{53},
Ryo Kandori\altaffilmark{6},
Jeremy Kasdin\altaffilmark{51},
Gillian R. Knapp\altaffilmark{13},
Tomoyuki Kudo\altaffilmark{5},
Nobuhiko Kusakabe\altaffilmark{6},
Masayuki Kuzuhara\altaffilmark{47,5},
Jungmi Kwon\altaffilmark{15},
Julien Lozi\altaffilmark{5},
Frantz Martinache\altaffilmark{52},
Taro Matsuo\altaffilmark{16},
Satoshi Mayama\altaffilmark{17},
Michael W. McElwain\altaffilmark{2},
Shoken Miyama\altaffilmark{18},
Jun-Ichi Morino\altaffilmark{6},
Amaya Moro-Martin\altaffilmark{38,39},
Takao Nakagawa\altaffilmark{41},
Tetsuo Nishimura\altaffilmark{5},
Tae-Soo Pyo\altaffilmark{5},
Eugene Serabyn\altaffilmark{20},
Hiroshi Suto\altaffilmark{6},
Ray W. Russel \altaffilmark{45},
Ryuji Suzuki\altaffilmark{6},
Michihiro Takami\altaffilmark{22},
Naruhisa Takato\altaffilmark{5},
Hiroshi Terada\altaffilmark{5},
Christian Thalmann\altaffilmark{23},
Edwin L. Turner\altaffilmark{9},
Taichi Uyama\altaffilmark{15},
Kevin R. Wagner\altaffilmark{49},
Makoto Watanabe\altaffilmark{25},
Toru Yamada\altaffilmark{26},
Hideki Takami\altaffilmark{6},
Tomonori Usuda\altaffilmark{15},
Motohide Tamura\altaffilmark{6,15}}
\altaffiltext{1}{Homer L. Dodge Department of Physics, University of Oklahoma, Norman, OK 73071, USA; erich66210@ou.edu, wisniewski@ou.edu}
\altaffiltext{2}{Exoplanets and Stellar Astrophysics Laboratory, Code 667, Goddard Space Flight Center, Greenbelt, MD
20771, USA}
\altaffiltext{3}{Eureka Scientific, 2452 Delmer, Suite 100, Oakland CA 96002, USA}
\altaffiltext{4}{Goddard Center for Astrobiology}
\altaffiltext{5}{National Astronomical Observatory of Japan, Subaru Telescope, National Institutes of Natural Sciences, 650 North A‘ohok¯u Place, Hilo, HI 96720, U.S.A}
\altaffiltext{6}{National Astronomical Observatory of Japan, 2-21-1, Osawa, Mitaka, Tokyo, 181-8588, Japan}
\altaffiltext{7}{Laboratoire Lagrange (UMR 7293), Universite de Nice-Sophia Antipolis, CNRS, Observatoire de la Cote d'Azur, 28 avenue Valrose, F-06108 Nice Cedex 2, France}
\altaffiltext{8}{Max Planck Institute for Astronomy, K\"{o}nigstuhl 17, D-69117 Heidelberg, Germany}
\altaffiltext{9}{Astrophysics Department, Institute for Advanced Study, Princeton, NJ 08540, USA}
\altaffiltext{10}{Department of Physics and Astronomy, College of Charleston, 58 Coming St., Charleston, SC 29424, USA}
\altaffiltext{11}{Universit\"{a}ts-Sternwarte M\"{u}nchen, Ludwig-Maximilians-Universit\"{a}t, Scheinerstr. 1, D-81679 M\"{u}nchen, Germany}
\altaffiltext{12}{Institute for Astronomy, University of Hawaii, 640 N. A'ohoku Place, Hilo, HI 96720, USA}
\altaffiltext{13}{Department of Astrophysical Science, Princeton University, Peyton Hall, Ivy Lane, Princeton, NJ 08544, USA}
\altaffiltext{14}{Department of Earth and Planetary Sciences, Tokyo Institute of Technology, 2-12-1 Ookayama, Meguro-ku,
Tokyo 152-8551, Japan}
\altaffiltext{15}{Department of Astronomy, The University of Tokyo, 7-3-1, Hongo, Bunkyo-ku, Tokyo, 113-0033, Japan}
\altaffiltext{16}{Department of Astronomy, Kyoto University, Kitashirakawa-Oiwake-cho, Sakyo-ku, Kyoto 606-8502, Japan}
\altaffiltext{17}{The Center for the Promotion of Integrated Sciences, The Graduate University for Advanced Studies (SOKENDAI), Shonan International Village, Hayama-cho, Miura-gun, Kanagawa 240-0193, Japan}
\altaffiltext{18}{Hiroshima University, 1-3-2, Kagamiyama, Higashi-Hiroshima 739-8511, Japan}
\altaffiltext{20}{Jet Propulsion Laboratory, California Institute of Technology, Pasadena, CA, 91109, USA}
\altaffiltext{22}{Institute of Astronomy and Astrophysics, Academia Sinica
11F of Astronomy-Mathematics Building, AS/NTU
No.1, Sec. 4, Roosevelt Rd, Taipei 10617, Taiwan, R.O.C}
\altaffiltext{23}{Institute for Astronomy, ETH Zurich, Wolfgang-Pauli-Strasse 27, 8093 Zurich, Switzerland}
\altaffiltext{25}{Department of Cosmosciences, Hokkaido University, Kita-ku, Sapporo, Hokkaido 060-0810, Japan}
\altaffiltext{26}{Astronomical Institute, Tohoku University, Aoba-ku, Sendai, Miyagi 980-8578, Japan}
\altaffiltext{28}{Graduate School of Science, Osaka University, 1-1 Machikaneyama, Toyonaka, Osaka 560-0043, Japan}
\altaffiltext{32}{Department of Physics, University of Cincinnati, Cincinnati, OH 45221, USA}
\altaffiltext{33}{Space Science Institute, 475 Walnut Street, Suite 205, Boulder, CO 80301, USA}
\altaffiltext{37}{Department of Astronomy, Stockholm University, AlbaNova University Center, SE-10691 Stockholm, Sweden}
\altaffiltext{38}{Space Telescope Science Institute, 3700 San Martin Dr., Baltimore, MD 21218, USA}
\altaffiltext{39}{Center for Astrophysical Sciences, Johns Hopkins University, Baltimore, MD 21218, USA}
\altaffiltext{41}{Department of Space Astronomy and Astrophysics Institute of Space \& Astronautical Science (ISAS) Japan Aerospace Exploration Agency (JAXA) 3-1-1 Yoshinodai, Chuo-ku, Sagamihara, Kanagawa 252-5210, Japan}
\altaffiltext{45}{The Aerospace Corporation, Los Angeles, CA, USA} 
\altaffiltext{46}{NASA-Ames Research Center, Moffett Field, CA USA}
\altaffiltext{47}{Astrobiology Center of NINS, 2-26-1, Osawa, Mitaka, Tokyo, 231-8588, Japan}
\altaffiltext{48}{Astrobiology Center, National Institutes of Natural Sciences, 2-21-1 Osawa, Mitaka, Tokyo, Japan}
\altaffiltext{49}{Steward Observatory, University of Arizona, Tucson, AZ 85721, USA
}
\altaffiltext{50}{NASA Goddard Space Flight Center}
\altaffiltext{51}{Department of Mechanical and Aerospace Engineering Princeton University, Princeton, New Jersey 08544, USA}
\altaffiltext{52}{Observatoire de la Cote d’Azur, Boulevard de l’Observatoire, 06300 Nice, France}
\altaffiltext{53}{Department of Astronomy, California Institute of Technology, Pasadena, CA 91125, USA}
\altaffiltext{54}{Department of Physics, University of Notre Dame, 225 Nieuwland Science Hall, Notre Dame, IN 46556, USA}
\altaffiltext{55}{Department of Physics \& Astronomy, University of Victoria, Victoria, BC V8P 1A1, Canada}


\begin{abstract}
We present H-band polarized scattered light imagery and JHK high-contrast spectroscopy of the protoplanetary disk around HD 163296 observed with the HiCIAO and SCExAO/CHARIS instruments at Subaru Observatory.  The polarimetric imagery resolve a broken ring structure surrounding HD 163296 that peaks at a distance along the major axis of 0$\farcs$65 (66 au) and extends out to 0$\farcs$98 (100 AU) along the major axis.  Our 2011 H-band data exhibit clear axisymmetry, with the NW- and SE-side of the disk exhibiting similar intensities.  Our data are clearly different than 2016 epoch H-band observations from VLT/SPHERE that found a strong 2.7x asymmetry between the NW- and SE-side of the disk.  Collectively, these results indicate the presence of time variable, non-azimuthally symmetric illumination of the outer disk. While our SCExAO/CHARIS data are sensitive enough to recover the planet candidate identified from NIRC2 in the thermal IR, we fail to detect an object with JHK brightness nominally consistent with this object.  This suggests that the candidate is either fainter in JHK bands than model predictions, possibly due to extinction from the disk or atmospheric dust/clouds, or that it is an artifact of the dataset/data processing, such as a residual speckle or partially subtracted disk feature.  Assuming standard hot-start evolutionary models and a system age of 5 $Myr$, we set new, direct mass limits for the inner (outer) ALMA-predicted protoplanet candidate along the major (minor) disk axis of of 1.5 (2) $M_{J}$.
\end{abstract}


\section{Introduction}
Protoplanetary disks are dust and gas disks around young stars that guide the accretion of material onto forming stars and serve as the birthplace of planets. 
Direct imaging of protoplanetary disks reveals likely sites of active planet formation, may identify planets in the final stages of assembly (protoplanets), and probes the interaction between protoplanets and the disk material from which they form.
Herbig Ae/Be stars \citep{herbig1960}, the intermediate mass analogs to T Tauri stars, are known to both host protoplanetary disks and often exhibit evidence of ejecting material via collimated, bi-polar jets \citep{herbig1950,grady2000,ellerbroek2014,bally2016}.
The protoplanetary disks around Herbig Ae/Be stars exhibit a variety of structures -- with some hosting spiral arms \citep{hashimoto2011} and others that are flat and settled causing self-shadowing of the disk \citep{meeus2001} -- and may host some of the first directly-imaged jovian protoplanets \citep{Quanz2013,Currie2015a}.

HD 163296 is a young ($5.1^{+0.3}_{-0.8}$ Myr old \citealt{montesinos2009} to $7.6^{+1.1}_{-1.2}$ \citealt{vioque2018}) Herbig Ae protoplanetary disk system located at a distance of 101.5 $\pm$ 1.2 $pc$ \citep{gaia2016,gaia2018}. 
The disk has been spatially resolved by ground- and space-based observing platforms at a multitude of wavelengths, including: optical (HST/STIS: \citealt{grady2000}, HST/ACS \citealt{wisniewski2008}), near-infrared (IR) (VLT/NACO: \citealt{garufi2014,garufi2017}, Gemini/GPI: \citealt{monnier2017}, VLT/SPHERE: \citealt{muro2018}, Subaru/CIAO: \citealt{fukagawa2010}, Keck/NIRC2: \cite{guidi2018}), and radio wavelengths (VLA: \citealt{guidi2016}, ALMA: \citealt{guidi2016,isella2016}).

Spatially-resolved imaging observations have revealed a complex circumstellar environment and evidence for active planet formation at wide separations around HD 163296.  
Its disk extends to at least to 4$\farcs$4 (447 AU) in optical scattered light \citep{wisniewski2008}.  While near-IR observations reveal a 64 AU-scale inner dust ring \citep{garufi2014,garufi2017,monnier2017,muro2018}, 1.3 mm continuum ALMA imaging \citep{isella2016} revealed three azimuthal gaps in the disk located at 0$\farcs$49, 0$\farcs$82, and 1$\farcs$31 (50, 83, and 133 au respectively given GAIA-DR2 distance of 101.5 pc).  The surface distribution of small dust grains in the outer disk appears low, owing to settling or partial-to-complete depletion \citep{muro2018}.
Keck/NIRC2 thermal infrared imaging led to the discovery of a candidate 7 $M_{\rm J}$ protoplanet just exterior to the inner ring \citep{guidi2018}, while modeling of ALMA gas emission 
data suggest Jovian planets at 83 and 137 au \citep{teague2018} and/or a single Jovian on an even wider orbit \citep[260 au][]{pinte2018}.

Multi-epoch observations have revealed a wealth of variability in the HD 163296 system likely traceable to dynamical processes in the inner disk region.  Both IR spectra and visibilities from optical inteferometry show variability possibly connected to changes in the inner disk or the system's wind component \citep{sitko2008,tan2008}.   Long-term optical photometric and IR spectroscopic monitoring revealed suggestive evidence of a 16 year periodicity, with optical fluxes dimming when the IR fluxes reach a maximum level \citep{ellerbroek2014,sitko2008}, on similar timescales as the ejection of Herbig-Haro objects \citep{ellerbroek2014}.  
The star's accretion rate increased over 1 dex over $\sim$ 15 years \citep{mendigut2013}. However, no clear correlation between these variations and the 16 year optical infrared periodicity has yet been found.
CO ro-vibrational emission lines exhibit variability possibly connected to changes in the disk wind or episodic accretion \citep{heinbertelsen2016}.

Spatially-resolved imaging may also reveal evidence for variability -- time-dependent changes in the disk's surface brightness and morphology potentially linked to variable illumination \citep{wisniewski2008}.
However, despite this plethora of variability observed, the lack of contemporaneous observations of both the inner and outer regions of the HD 163296 disk limits efforts to connect these phenomenon to one another.
  
In this paper, we present multi-epoch near-infrared scattered light imaging of HD 163296, obtained at $H$-band in polarized light as part of the Strategic Exploration of Exoplanets and Disks with Subaru (SEEDS) survey \citep{tumara2009} and in total intensity in $JHK$ using \textit{Subaru Coronagraphic Extreme Adaptive Optics} (SCExAO) \citep{Jovanovic2015a} coupled with the CHARIS integral field spectrograph (Section \ref{sec:obs}). To help parse and complement these data probing the outer disk, we acquired near-contemporaneous IR spectra to characterize the inner disk region of the system. We modeled the H-band scattered light images and near-IR spectra using a well-established 3D Monte Carlo Radiative Transfer code to create a more coherent, full picture of the system at this epoch (Section \ref{sec:analysis}). Finally we discuss the implications of our results in Section \ref{sec:discussion} including deeper constrains on protoplanets around HD 163296 with the new SCExAO/CHARIS data. 

\section{Observations and Data Reduction}\label{sec:obs}
\subsection{HiCIAO Imagery}\label{sec:hiciao_obs}
We obtained high contrast H-band imaging of HD 163296 using the HiCIAO instrument \citep{hodapp2008} along with the AO-188 system \citep{hayano2008,hayano2010} at the Subaru Observatory on 2011 August 3.  We 
used a circular occulting mask having a diameter of 0$\farcs$3, and observed the system in 
standard Polarized Differential Imaging (sPDI) mode at four wave-plate positions (0$^\circ$, 22$^\circ$.5, 45$^\circ$, 67$^\circ$.5).  We obtained 72 frames using 30 second exposures, yielding a total of 18 complete wave-plate sets. We determined that 8 wave-plate sets had lower AO performance, and discarded them during the reduction of the data. We also obtained a short, direct H-band photometric observation of HD 163296, and determined that the source's brightness at this epoch was 5.62 $\pm$ 0.05 mag.

We reduced our observations using standard double differencing techniques, as described in \citet{hashimoto2011}. To briefly summarize, the two sub-images of each frame contain an ordinary and an extra-ordinary image, which can be summed and subtracted from their 90$^\circ$ counterparts to create stokes parameter $-$Q, +Q, $-$U, and +U images.
The Q and U frames were then rotated into a common orientation, corrected for instrumental polarization, and summed to create final Q and U images.  We corrected these data for the presence of a residual polarized halo having the properties of p = 1.00 $\pm$ 0.05\% and $\theta$ = 42.5 $\pm$ 1.5$^\circ$.  Final polarized intensity (PI) imagery was created from the total Q and U data, using $PI = \sqrt[]{Q^2 + U^2}$, as shown in Figure \ref{fig:images}.

To further simplify the analysis of our imagery, we adopt the now common practice of assuming single scattering, and rotated all of the light that is polarized perpendicular to the star by the angle $\phi$ into a Q$_\phi$ image and all of the light that is polarized parallel to the star into a U$_\phi$ image as defined below \citep{schmid2006}. 

\begin{align}
Q_\phi = Q \times \cos 2 \phi + U \times \sin 2 \phi \\
U_\phi = Q \times \sin 2 \phi + U \times \cos 2 \phi
\end{align} \\

The final Q$_\phi$ and U$_\phi$ imagery for HD 163296 are shown in Figure \ref{fig:images}.  Little coherent signal appears present in the U$_\phi$ image, which helps confirm that little residual instrumental contaminants remain in these data.  Next, we computed a signal to noise (SN) image, following the procedure outlined by \citet{ohta2016}. In summary, we computed the noise by measuring the standard deviation of every pixel in each of the Q and U frames used to construct the final imagery, then divided by the square root of the number of frames. The resultant SN image is shown in Figure \ref{fig:images}.

\subsection{Near-Infrared Spectra from SpeX, BASS, and TripleSpec} \label{sec:nearIRspec}
We also observed HD 163296 multiple times with several near-IR instruments on NASA's Infrared Telescope Facility (IRTF) and at Apache Point Observatory (APO).  We observed HD 163296 using the SpeX spectrograph \citep{rayner2003} at IRTF in its short-wavelength mode (0.8 - 2.4 $\mu m$) and long-wavelength mode (2.3-5.5 $\mu m$) on 2011 July 31, 2016 May 4, and 2018 June 24. These observations are contemporaneous with the HiCAIO 2011 observation (Section \ref{sec:hiciao_obs}), the Gemini/GPI observation (Section \ref{sec:epoch}), and the second SCExAO/CHARIS observations (Section \ref{sec:charis_obs}) respectively.
We observed HD 163296 using the TripleSpec spectrograph \citep{wilson2004} at the APO 3.5m telescope, covering a spectral range of (0.95 - 2.46 $\mu m$), on 2018 May 16. This observation is contemporaneous with the first SCExAO/CHARIS observation (Section \ref{sec:charis_obs}).
We observed the nearby A0V star HD 163336 to perform telluric corrections for both the SpeX and TripleSpec observations.  These data were reduced and calibrated using the standard reduction packages \textit{Spextool} and \textit{Triplespectool} \citep{vacca2003,cushing2004}.  We also observed HD 163296 with The Aerospace Corporation's Broad-band Array Spectrograph System (BASS), which covers two wavelength bands from 2.9-6 $\mu m$ and 6-13.5 $\mu m$ respectively, on 2011 August 1.  HD 163336 was observed with BASS to flux calibrate these data.  The instrument and data reduction method are fully described in \citet{wagner2015}. 
These SpeX, TripleSpec, and BASS spectra are plotted in Figure \ref{fig:spex}.

\subsection{SCExAO/CHARIS High-Contrast Near-Infrared Spectroscopy} \label{sec:charis_obs}
We observed HD 163296 on 2018 May 22 and 2018 July 1 at the Subaru Observatory with SCExAO coupled with the CHARIS integral field spectrograph operating in low-resolution ($R\sim 20$), 
broadband (1.13--2.39 $\micron$) mode, covering the $JHK$ filters simultaneously \citep{Groff2015}. For the May observations, the conditions were stable with 0\farcs{}4 seeing and 6--7 m s$^{-1}$ winds. Our observations consisted of co-added 60.4-second frames totaling $\sim$30 minutes of integration time and covering a modest parallactic angle rotation ($\Delta$PA = 14.8$^{o}$).   Due to highly variable conditions for the July observations, we obtained shorter exposures (30.9 $s$) and removed roughly 50\% of the frames with poor AO correction, yielding $\sim$ 40 minutes of data covering 30.9$^{o}$ of parallactic angle motion\footnote{While a real-time estimate of the Strehl ratio (S.R.) was not recorded for these data sets, the raw contrast for the May data was just slightly poorer than that obtained for $\kappa$ And observations achieving S.R. $\sim$ 0.90--0.92 in $H$ band \citep{Currie2018b}.   Raw contrasts for the July data considered in our study are roughly a factor of 2.5--3 worse at 0\farcs{}4, more characteristic of performance at S.R. $\sim$ 0.65--0.70.}.

We followed the standard setup used for SCExAO/CHARIS broadband observations \citep{Currie2018b,Goebel2018}, using the Lyot coronagraph with the 217 mas occulting spot and bracketing our coronagraphic sequence with blank sky frames to remove sky emission and instrumental artifacts.   We used satellite spots produced from a 25 nm modulation on SCExAO's deformable mirror for spectrophotometric calibration and image registration \citep{Jovanovic2015b}.   For data cube extraction, we utilized the least-squares algorithm from the CHARIS Data Reduction Pipeline \citep{Brandt2017}.   Basic data processing, including sky subtraction, image registration, etc., follows methods used for recent SCExAO/CHARIS broadband studies \citep{Currie2018a,Currie2018b,Goebel2018}.  

Spectrophotometrically calibrating CHARIS data for pre-transitional disk sources like HD 163296 require either observations of a separate spectral standard or contemporaneous near-IR spectra. We opt for the latter, using the IRTF/SpeX and APO/Triplespec data previously discussed in Section \ref{sec:nearIRspec}. The spectra show only minor differences between epochs.      

We explored a range of point-spread function (PSF) subtraction approaches leveraging on \textit{angular differential imaging} \citep[ADI][]{Marois2006}, \textit{spectral differential imaging} \citep[SDI][]{SparksFord2002}, and combinations of the two \citep[ASDI, e.g.][]{Marois2014}.   We further considered a variety of PSF subtraction algorithms, including A-LOCI \citep{Currie2012,Currie2018b}, KLIP \citep{Soummer2012}, and classical PSF subtraction \citep{Marois2006}.   The approach implemented for $\kappa$ And in \citet{Currie2018b}, using A-LOCI to subtract the PSF in ADI and then again to remove  residuals in SDI mode, yielded the best speckle suppression while preserving the signal from the disk.   Due to the limited parallactic angle motion of both data sets (especially in May) and the presence of the disk, we utilized large optimization zones for the ADI step, employed local masking in the SDI step, and imposed a rotation/magnification criterion of $\delta$ = 0.5--1.0 PSF footprints in both steps to construct a reference PSF \citep[see ][]{Lafreniere2007}.   For both steps, we used a \textit{singular value decomposition} (SVD) cutoff of 10$^{-6}$ to solve the set of linear equations that result in the weighted reference PSF for each region of each data cube slice \citep[see ][]{Currie2015a}.

Figure \ref{fig:charis} shows broadband (wavelength-collapsed) CHARIS images of HD 163296 from SCExAO/CHARIS for the May (left) and July (right) epochs after removing the stellar PSF through both ADI and SDI.   Despite poor field rotation (May data) or variable conditions (July data), we clearly detect  the outer ring of emission seen in polarimetry, which appears as a sharply-defined crescent defining the forward-scattering edge of the structure.   Self-subtraction footprints due to both ADI and SDI flank the ring.   In individual passbands, the disk is just marginally visible in $J$ band but is well separated from residual speckle noise in $H$ and $K$.   

We defined a conservative lower limit to the signal-to-noise (SNR) ratio of the trace of the disk in broadband, adopting the standard practice of replacing each pixel by the sum within its aperture, defining a radial-dependent noise profile, and applying a finite-element correction for the noise \citep{Currie2011a,Mawet2014}.  To be conservative, we include signal from the disk in our estimate of the noise profile.  Except at the semi-minor axis, where the disk signal is attenuated by self-subtraction, the disk trace is decisively detected, with a SNR per resolution element ranging from 3 to 8.5.  

Our data do not reveal the candidate protoplanet identified in Keck/NIRC2 $L_{\rm p}$ data from \citet{guidi2018} nor the companions predicted from ALMA data \citep{teague2018}.   The inner disk seen by GPI polarimetry \citep{monnier2017} is also not visible, likely due to heavy self-subtraction due to poor field rotation.   The position of the \citeauthor{guidi2018} candidate lies well separated from the ring and residual speckle noise; the SNR maps show no convolved pixel within one PSF footprint ($\sim$ 0\farcs{}08) of this position with a significance greater than 1.3$\sigma$.   More conservative reductions (e.g. larger rotation gap; higher SVD cutoff) may show slightly elevated residual emission consistent with additional extended structure at this separation (e.g . additional ring material).  However, this signal is not statistically significant and is simpler to explain as residual speckle noise instead.

\section{Analysis of the H-band Polarimetry Data}\label{sec:analysis}

In this section, we characterize the distribution of scattered light in our H-band imagery, and construct a Monte Carlo Radiative Transfer (MCRT) model to help interpret the contemporaneous H-band scattered light imagery and near-IR spectra.

\subsection{Geometry of the Disk}
\begin{figure}
\centering
\includegraphics[width=\columnwidth]{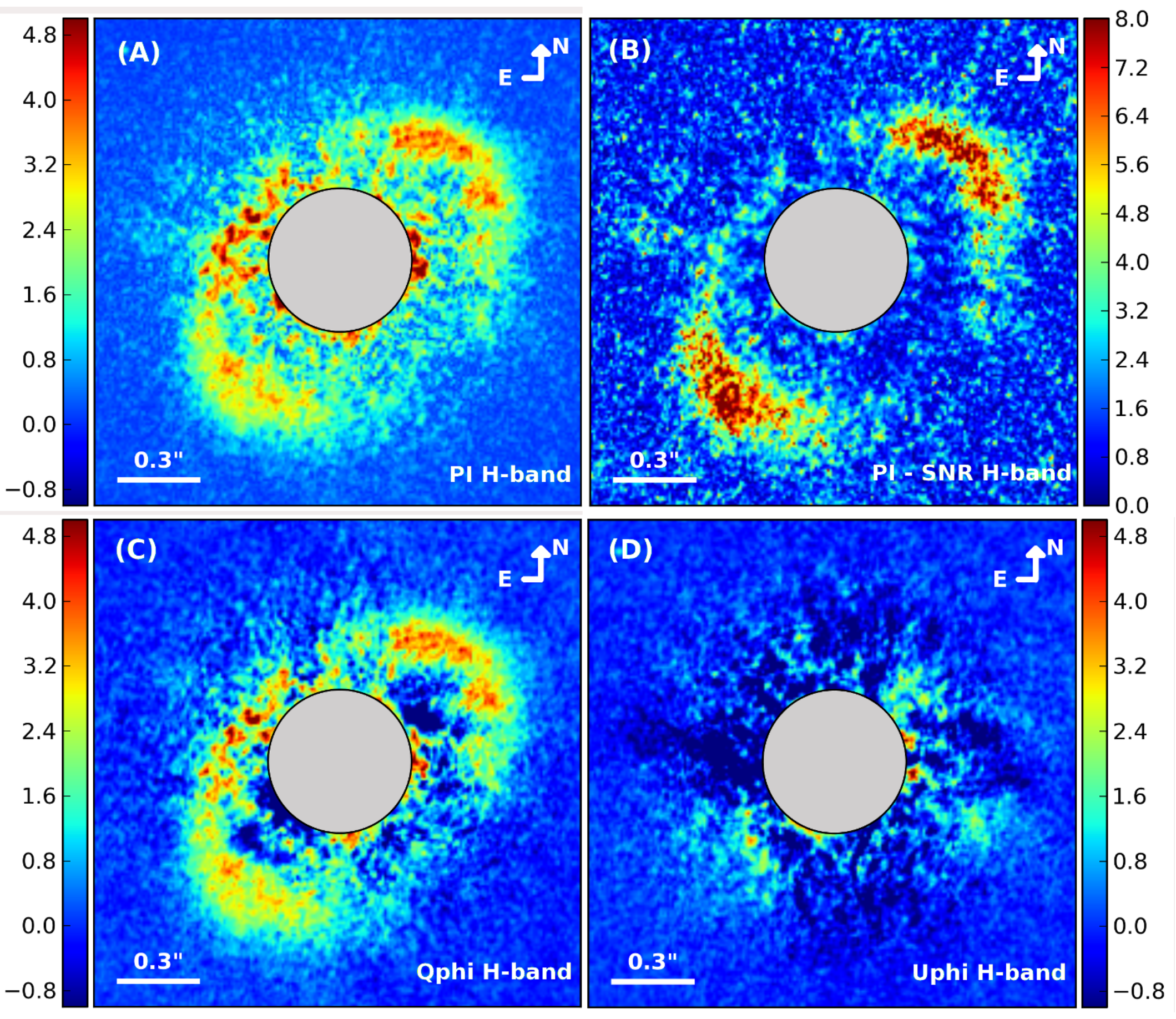}
\caption{H-band scattered light from the HD 163296 disk is clearly seen in 
polarized intensity (PI) (panel A), the SN map (panel B), and in Q$_{phi}$ imagery 
(panel C).  Little coherent signal is seen in the U$_{phi}$ image (panel D), 
indicating that these data are largely free from PSF residuals.  The PI (panel A), Q$_{phi}$ (panel C), and U$_{phi}$ (panel D) images are displayed on a linear scale with units of mJy, and have not been filtered.  We have applied a software mask having a radial size of 0$\farcs$3 (gray circles) to match the effective inner working angle of these data. For all panels, North is up and East is to the left. The Q and U images shown in panels C and D of this figure is available as the Data behind the Figure.}
\label{fig:images}
\end{figure}

\begin{figure}
\centering
\includegraphics[width=\columnwidth]{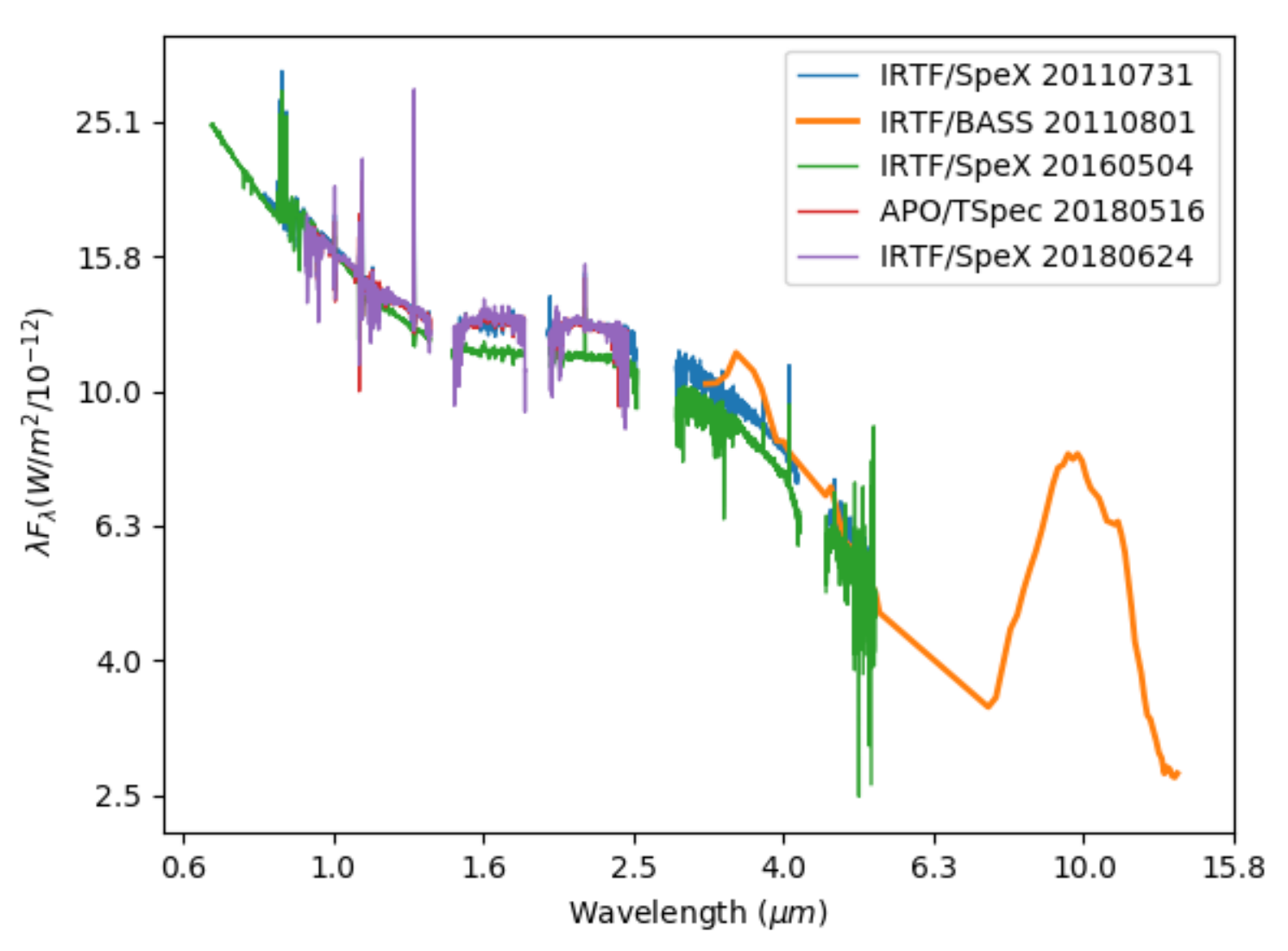}
\caption{5 epochs of flux calibrated IR spectra of HD 163296, taken with IRTF/SpeX, IRTF/BASS, or APO/TripeSpec, are shown. A full description of these observations can be found in Section \ref{sec:nearIRspec}. The spectra are plotted in log-log space.}
\label{fig:spex}
\end{figure}

\begin{figure}
\centering
\includegraphics[width=0.49\columnwidth]{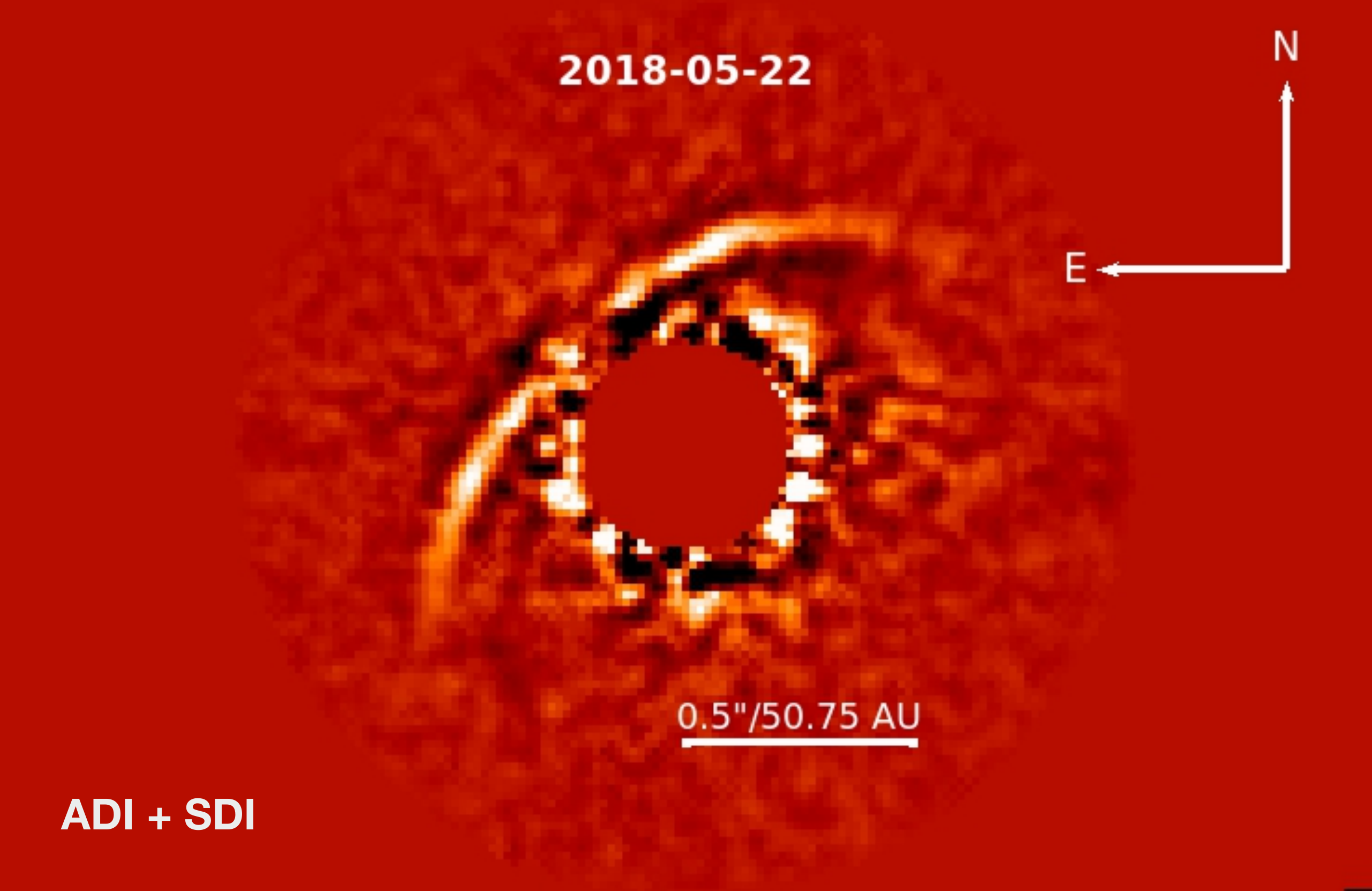}
\includegraphics[width=0.49\columnwidth]{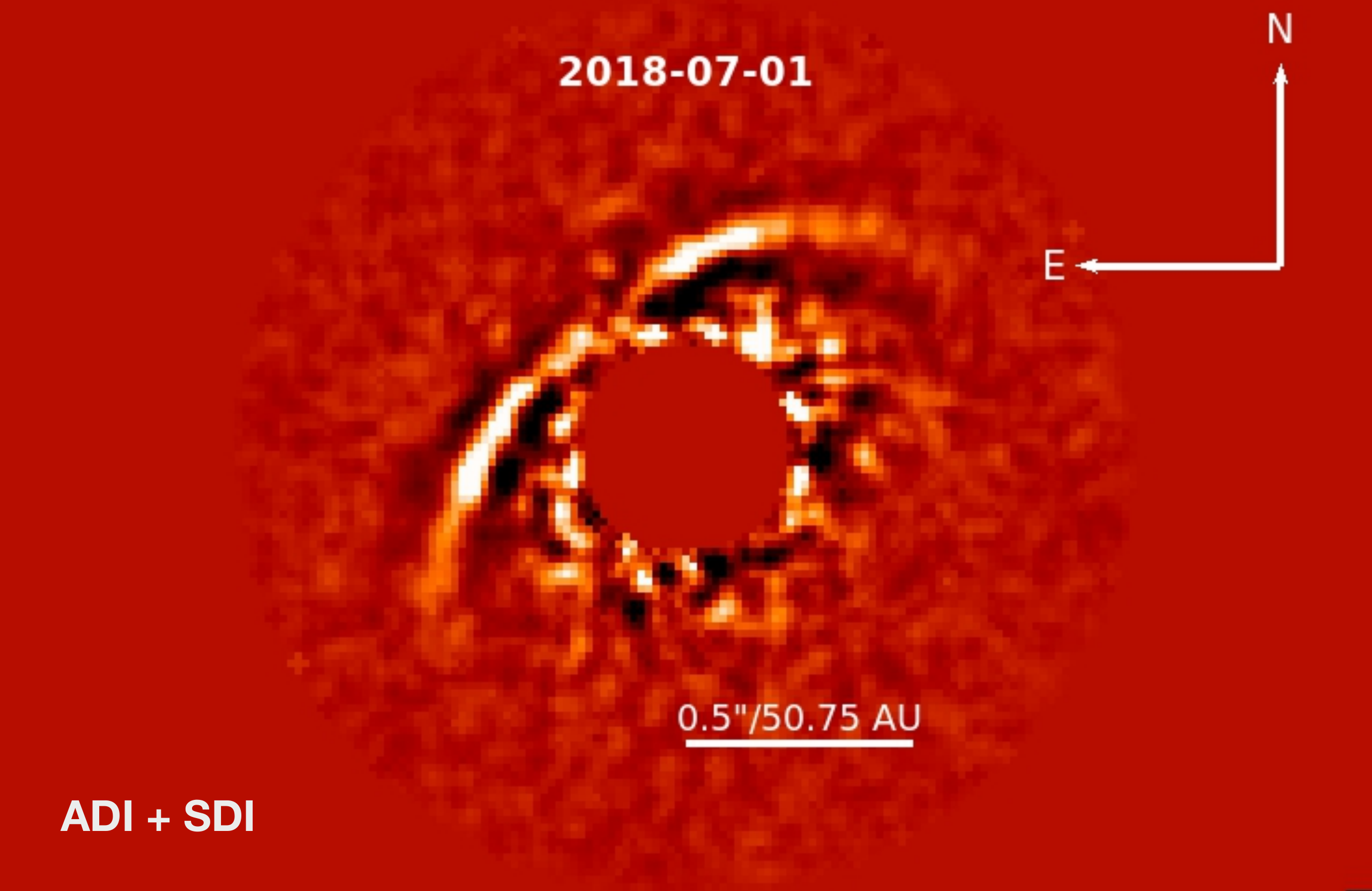}
\caption{SCExAO/CHARIS broadband (wavelength-collapsed) images from 2018 May (left) and 2018 July (right) after removing the stellar PSF through both ADI and SDI: the color scaling for both panels goes from -30 to 30 mJy arcsec$^{-2}$. In both data sets, self-subtraction footprints (dark regions) flank the disk signal, which is reduced due to processing. The throughput of the disk is slightly higher in the July data due to better field rotation; regions surrounding the disk show slightly less residual speckle noise in the May data due to better AO performance.}
\label{fig:charis}
\end{figure}

\begin{figure}
\centering
\includegraphics[width=\columnwidth]{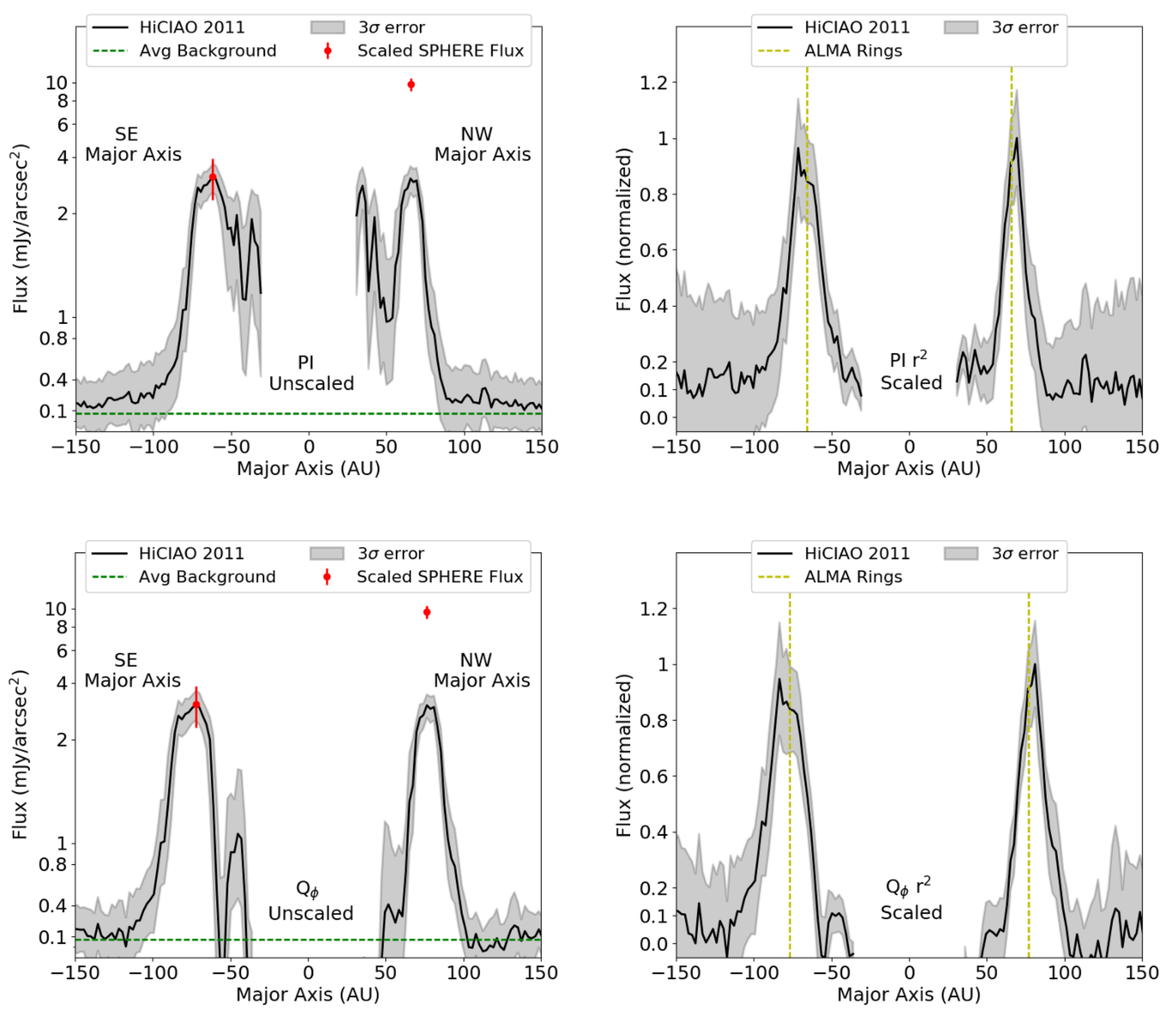}
\caption{Crosscuts along the major axis of the 2011 H-band PI image (top row) and Q$_{\phi}$ image (bottom row). The right column is the PI and Q$_{phi}$ images unscaled, and the left column is the PI and Q$_{\phi}$ with a $r^2$ scaling applied. The gray shaded area represents 3-$\sigma$ error bars. The red point is the scaled flux from the 2016 VLT/SPHERE observation reported by \citet{muro2018}.}
\label{fig:crosscuts}
\end{figure}

Figure \ref{fig:images} reveals the clear detection of scattered light surrounding the HD 163296 disk in our H-band imagery outside of the inner working angle of these data, 0$\farcs$3 (30.5 au).  The scattered light imagery reveals a broken ring structure that peaks at a distance along the major axis of 0$\farcs$65 (66 au) and extends out to 0$\farcs$98 (100 AU) along the major axis (see Figure \ref{fig:crosscuts}). Both the Q$_\phi$ and SN imagery exhibit little coherent signal between our inner working angle and the inner edge of the ring structure. 
We do not detect the inner disk component as previously detected by \citet{monnier2017} due to our larger inner working angle.
We therefore conclude that  
the small amount of scattered light interior to the ring in the PI image (panel A; Figure \ref{fig:images}) could arise from a mixture of residual, uncorrected flux from the PSF and scattered flux from an inner disk component that is within our masked region (see e.g. \citealt{takami2018}). The NE-side of the disk is known to be the near side \citep{rosenfeld2013} and the IR scattered light disk exhibits evidence of strong forward scattering \citep{guidi2018}.  The broken ring structure we observe is missing polarized intensity originating from the far-side of the disk (SW region, along the minor axis; see Figure \ref{fig:images}).   

We fit an ellipse to the scattered light ring using a least squares fitting code written by Ben Hammel and Nick Sullivan-Molina \footnote{https://github.com/bdhammel/least-squares-ellipse-fitting}, assuming the ring is a perfect circle projected at inclination.  Since a known bias of the code is to prefer a smaller ellipse by preferably fitting the inner points \citep{halif1998}, we choose to fit the peaks of the ring to mitigate this effect.  Due to the low signal along the SW minor axis and sporadic structure along the NE minor axis, we did not keep any vertical cuts between 70$^\circ$ $<$ PA $<$ 180$^\circ$ and between 270$^\circ$ $<$ PA $<$ 370$^\circ$.
We fit a gaussian to each vertical crosscut, producing the peak x,y position of the ring, and input these positions into the ellipse code described above.

\begin{figure}
\centering
\includegraphics[width=\columnwidth]{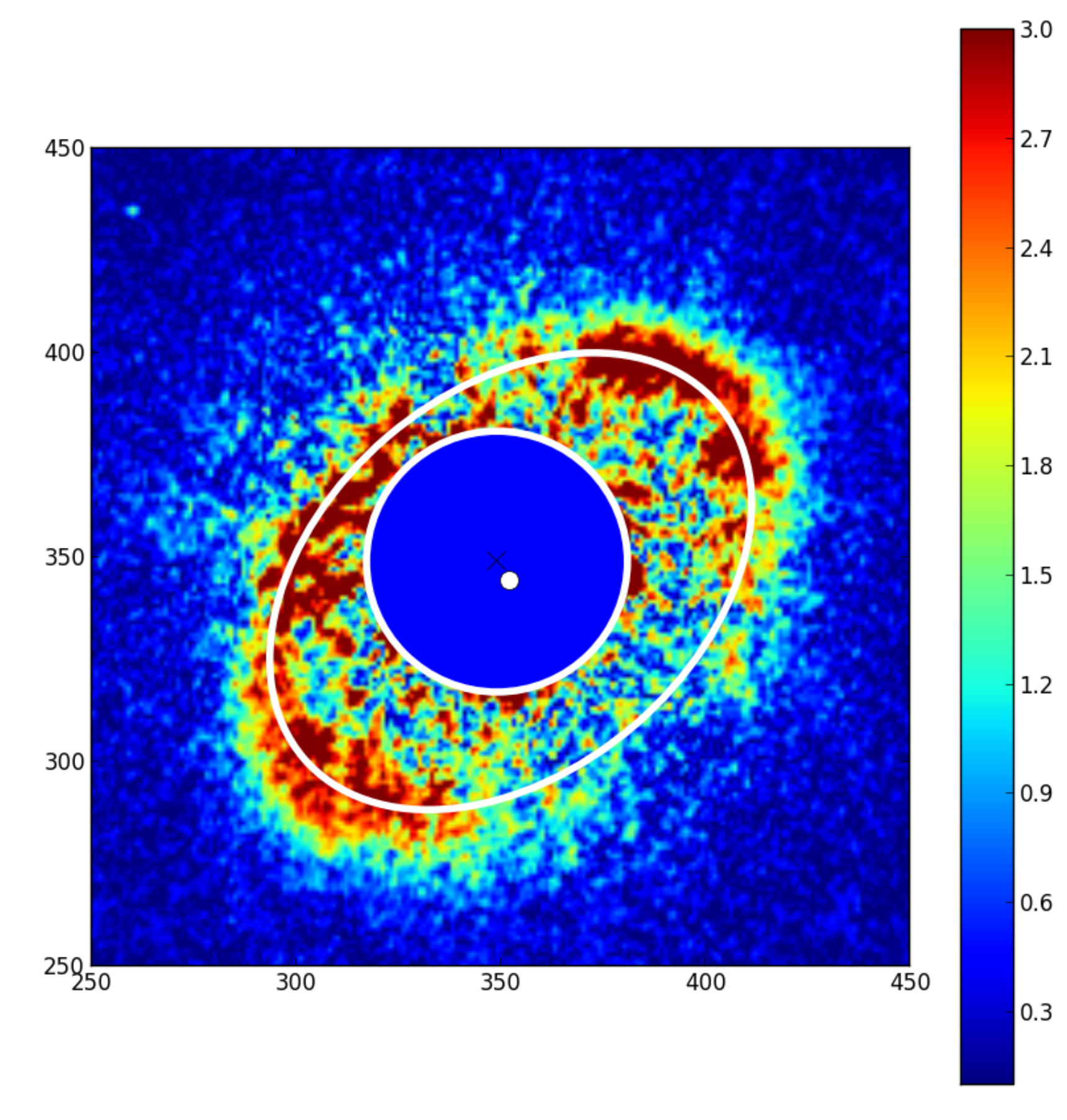}
\caption{Result of the best fit ellipse to our H-band PI data, where the central white dot is the center of the ellipse, the white ellipse is the peak of the ellipse, the black x marks the location of the star, and the blue circle marks the inner working angle. The ellipse was fit to the peak points along the main elliptical ring by fitting gaussians to the cross cuts along the ring. The best elliptical fit finds a minor axis offset of -0$\farcs$055. This value is consistent with those reported by \citet{garufi2014,monnier2017,muro2018} given their quoted 
uncertainties.}
\label{fig:ellipse}
\end{figure}

In order to estimate the error of our ellipse fit, we performed a Monte Carlo routine by randomly sampling the gaussian xy-coordinate errors and adding them to the xy-coordinates found above. We additionally applied a random rotation of the image between 0$^\circ$ and 1$^\circ$ to constrain the error associated with the interpolation of the image due to rotation.
We performed 500 iterations and used the average values of the 500 iterations as the best fit ellipse. The errors were estimated by taking the standard deviation of the parameters found with the 500 iterations. The best fit results are shown in Table \ref{tbl:fit}. The best fit ellipse is compared to the PI image in Figure \ref{fig:ellipse} shown as the white oval along with the center of the disk (small white circle) and the center of the star. 

Our measured inclination of the disk (41.4 $\pm$ 0.3$^{\circ}$) and PA (132.2$^\circ$ $\pm$ 0.3$^\circ$) is in agreement with the values derived from ALMA data of 42 $^\circ$ and 132$^\circ$ respectively \citep{isella2016}. Additionally, the offset of the minor axis from the central star that we find (-0$\farcs$0432 $\pm$ 0$\farcs$0016) is consistent with previous measurements, given their quoted errors when available (0$\farcs$06, \citealt{garufi2014}; 0$\farcs$105 $\pm$ 0$\farcs$045,  \citealt{muro2018}; 0$\farcs$1, \citealt{monnier2017}). 

\begin{figure}
\centering
\includegraphics[width=\columnwidth]{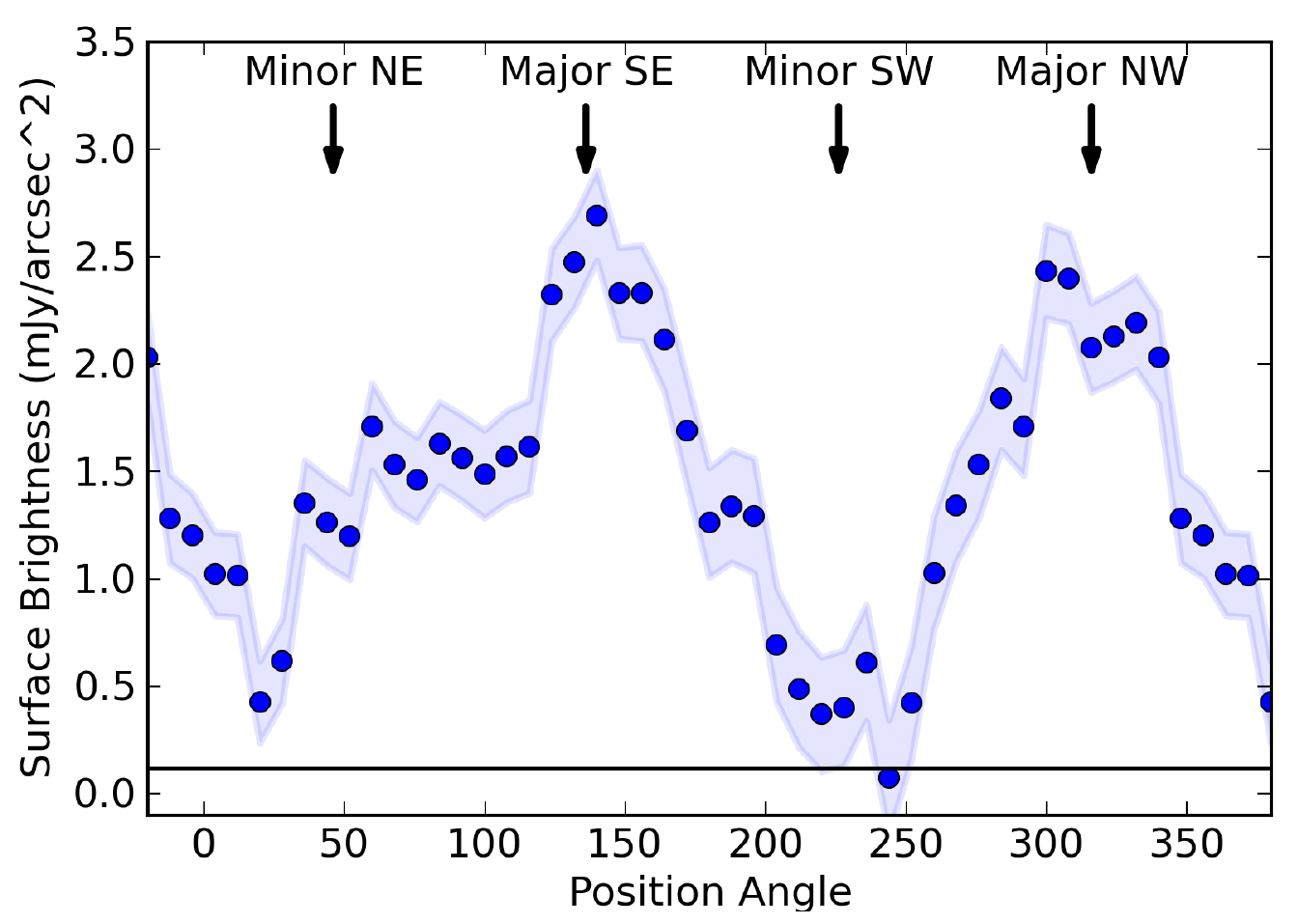}
\caption{Binned flux along the azimuthal ring located at 65 AU. Each bin is 8$^\circ$ wide and extends from a projected distance of 55 to 71 AU annulus along the ring seen in this figure.}
\label{fig:theta_flux}
\end{figure}

We applied an $r^2$ illumination correction to our data to better investigate the physical distribution of dust in the ring seen in Figure \ref{fig:images}. We then azimuthally binned the average flux per area of the ring between two concentric ellipses.  We adopted an inclination of 42$^\circ$, and constructed each bin to be $8^\circ$ wide and spanned a projected radial distance of 0$\farcs$55 - 0$\farcs$71 (55 - 72.5 au), to encompass the majority of the disk flux.  The binned disk flux is azimuthally symmetric along the major axis, with the NW- and SE-side of the disks exhibiting the same amount of polarized intensity (Figure \ref{fig:theta_flux}).  There is also a clear azimuthal asymmetry in the binned flux along the minor axis, with the near-side of the disk (NE-side) exhibiting substantially more flux than the far-side (SW-side).  We observe a deficit in scattered light flux along the near-side of the disk at a PA of 30$^\circ$ in both the binned imagery (Figure \ref{fig:theta_flux}) and  unbinned PI, Q$_{rot}$ images, SN map images (Figure \ref{fig:images}).  This feature coincides with the position angle of the disk brightness enhancement and the position angle of the candidate point source noted by \citet{guidi2018} and will be further discussed in Section \ref{sec:Lprime}.

\subsection{Modeling of the HD 163296 Disk}\label{sec:model}
To help interpret our imagery and contemporaneous IR spectroscopy of HD 163296, we utilized the 3D Monte Carlo Radiative Transfer code (MCRT), HOCHUNK3D \citep{whitney2013}. HOCHUNK3D allows the user to characterize the radial dust distribution, dust composition, and disk illumination parameters, and outputs a SED of the disk and imagery in a variety of user-defined bandpasses. The current version of HOCHUNK3D allows the user to decouple the disk into two dust distributions, allowing one to parameterize both a settled dust population towards the midplane and a different dust population in the upper surface layers of the disk.  These two dust populations can either be co-spatial, or have different radial sizes. The dust distribution of each disk is characterized by several power-law parameters: the radial power law ($\alpha$), the vertical gaussian distribution ($\beta$), and the height of the disk from the mid-plane ($h$).  Deviations from these power-laws such as a gap, spiral arms, warped disks, and walls can all be included.  The code also allows for the presence of a dusty envelope, which is parameterized by its minimum and maximum radius (R$_{min env}$, R$_{max env}$), and a dust density powerlaw (ENVEXP). The dusty envelope can also include gaps and a bipolar cavity. Following the techniques established by \citet{sitko2008,wagner2015,fernandes2018}, we use the dusty envelope as a proxy to model material ejected from the disk, aka a disk wind.

We constrained our model starting parameters by observations when possible, and adopted the parameters from Pikhartova et al. (in prep), who are using HOCHUNK3D to model the variations seen in two epochs of HD163296's SED, as a starting point for our parameter-space exploration.  ALMA observations of HD 163296 revealed the presence of 3 gaps located at 0$\farcs$49, 0$\farcs$82, and 1$\farcs$31 (50, 83, and 133 au respectively given GAIA-DR2 distance of 101.5 pc) \citep{isella2016}.  Since our HiCIAO imagery is only sensitive to the first dust ring and the near-IR SED is most sensitive to dust features closer to the star, 
we only include the inner gap in our model.  We allowed the two components of the dust distribution to be vertically stratified, and chose the radial extent of these distributions to match those observed for grains populating the midplane (250 au from VLA and ALMA observations; \citealt{guidi2016}) and surface layers (540 au, \citealt{isella2007,wisniewski2008}).  We note that while ALMA observations of the system were best described by a radial power-law multiplied by an exponential function \citep{isella2016}, HOCHUNK3D only uses a power-law function. 
Nevertheless, we did adjust the large grain dust distribution to match, as closely as possible, the dust distribution as measured by ALMA in the inner portion of the disk \citep{isella2016}.  The dust parameters for the large grain disk that we used are adopted from \citet{wood2002}, and are composed of amorphous carbon and silicon dust particles ranging in size up to 1 millimeter. The small grain disk and envelope dust parameters are from \citet{kim1994}, which is the average galactic ISM dust grain model.
  
We adopted an interstellar extinction of A$_V$ = 0 mag from \citet{ellerbroek2014}, who measured the level of extinction from the ejected HH-knots. Note that \citet{ellerbroek2014} concluded that the optical variability of the SED likely comes from on source reddening. In our model, we utilize the dusty envelope, a proxy to model disk wind, to replicate the on source reddening which is further discussed in section \ref{sec:time}.
We explored accretion rates ranging from $1.73 \times 10^{-7}$ to $4.35 \times 10^{-6}$ M$_{\odot}$, calculated from contemporaneous Br$\gamma$ emission line in the SpeX 2011 data, first presented in \citet{ellerbroek2014}, but adjusted for the new distance of 101.5 pc. 

We constrained these models using a SED (Figure \ref{fig:model_SED}) constructed from contemporaneous near-IR observations (Figure \ref{fig:spex}), along with non-contemporaneous photometry from the All WISE catalog \citep{wright2010}, 2MASS All Sky Survey \citep{cutri2003}, IRAS point source catalog \citep{helou1988}, and the historical variability of the V-band photometry as compiled in \citet{ellerbroek2014}.  We also constrained these models using the surface brightness profiles along the major axis of our HiCIAO H-band scattered imagery (Figure \ref{fig:crosscuts}).

We explored the parameter space of our models using a $\chi^2$ minimization scheme.  Namely, we calculated the $\chi^2$ for the SED fit, the surface brightness along the major axis, and the minor axis offset, and added these values in quadrature to find the total $\chi^2$ value. Since some of the SED data were not contemporaneous, we also calculated a separate $\chi^2$ value that only incorporated comparisons of contemporaneously obtained data to the model.  We began the iterative process with model runs of 5 million photons in order to find the best fit SED to the SpeX and Bass spectra.  Next, we increased the number of photons in each run to 50 million photons to obtain higher quality model H-band images, and convolved the model image with the PSF of the H-band image. We explored parameter space to produce the best fit $\chi^2$ value between model surface brightness along the major axis and the minor axis offset to the observed imagery. After finding the best chi-squared fit model image, we iteratively switched between the SED and the model until we found a model that optimized the combined chi-squared value, resulting in our best fit model.  We then re-ran this best fit model using 10$^{9}$ photons to produce the model SED and imagery used all of our figures. We remind readers that MCRT models, like HOCHUNK3D that employ a large family of parameters, suffer from parameter degeneracy, thus our best fit model is not unique \citep{dong2012}. 

\begin{figure}
\centering
\includegraphics[width=\columnwidth]{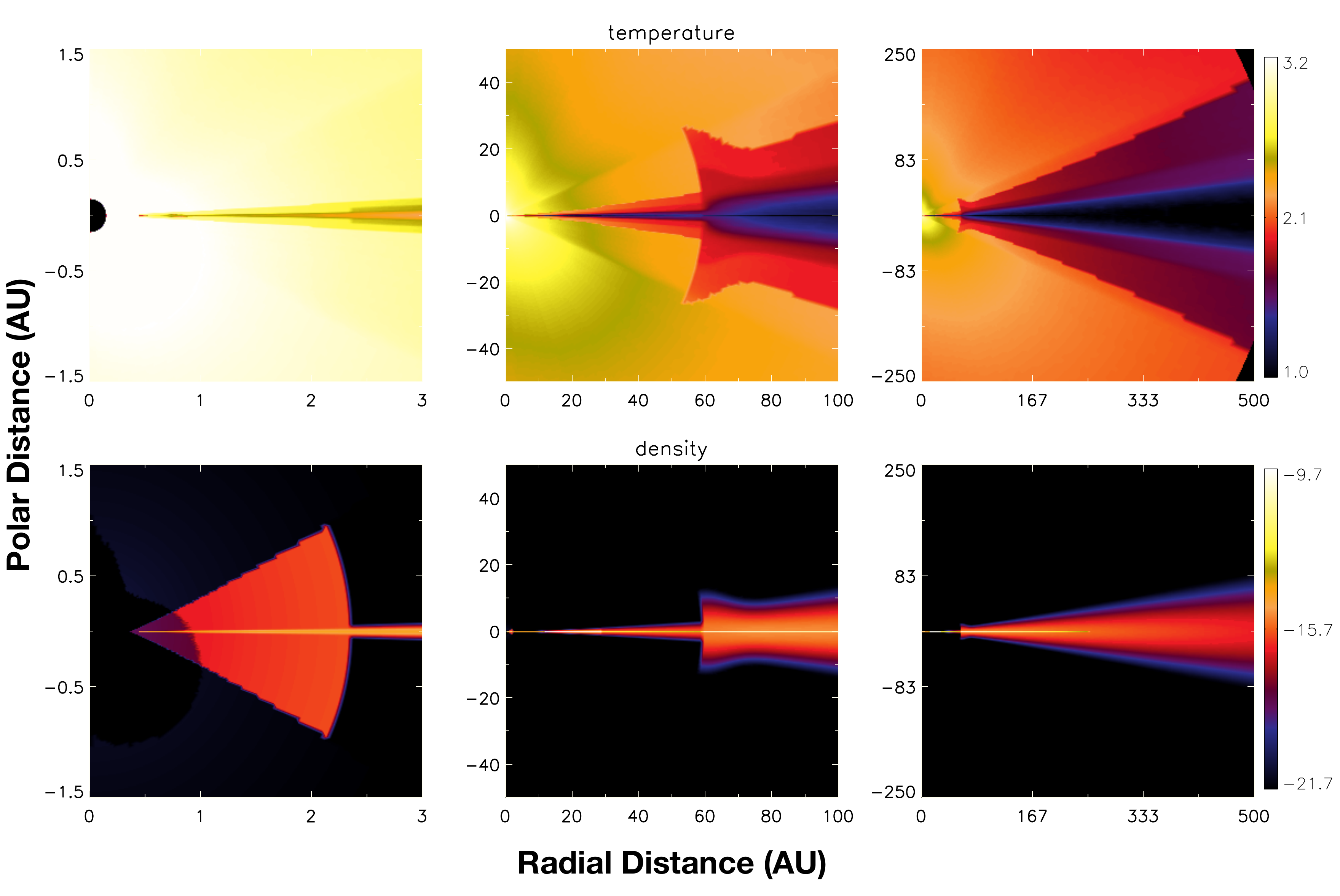}
\caption{The top row of panels present temperature profiles for three regions of our MCRT disk model. The bottom row of panels present the density profiles for these same three regions of the disk model.}
\label{fig:trho}
\end{figure}

\begin{figure}
\centering
\includegraphics[width=\columnwidth]{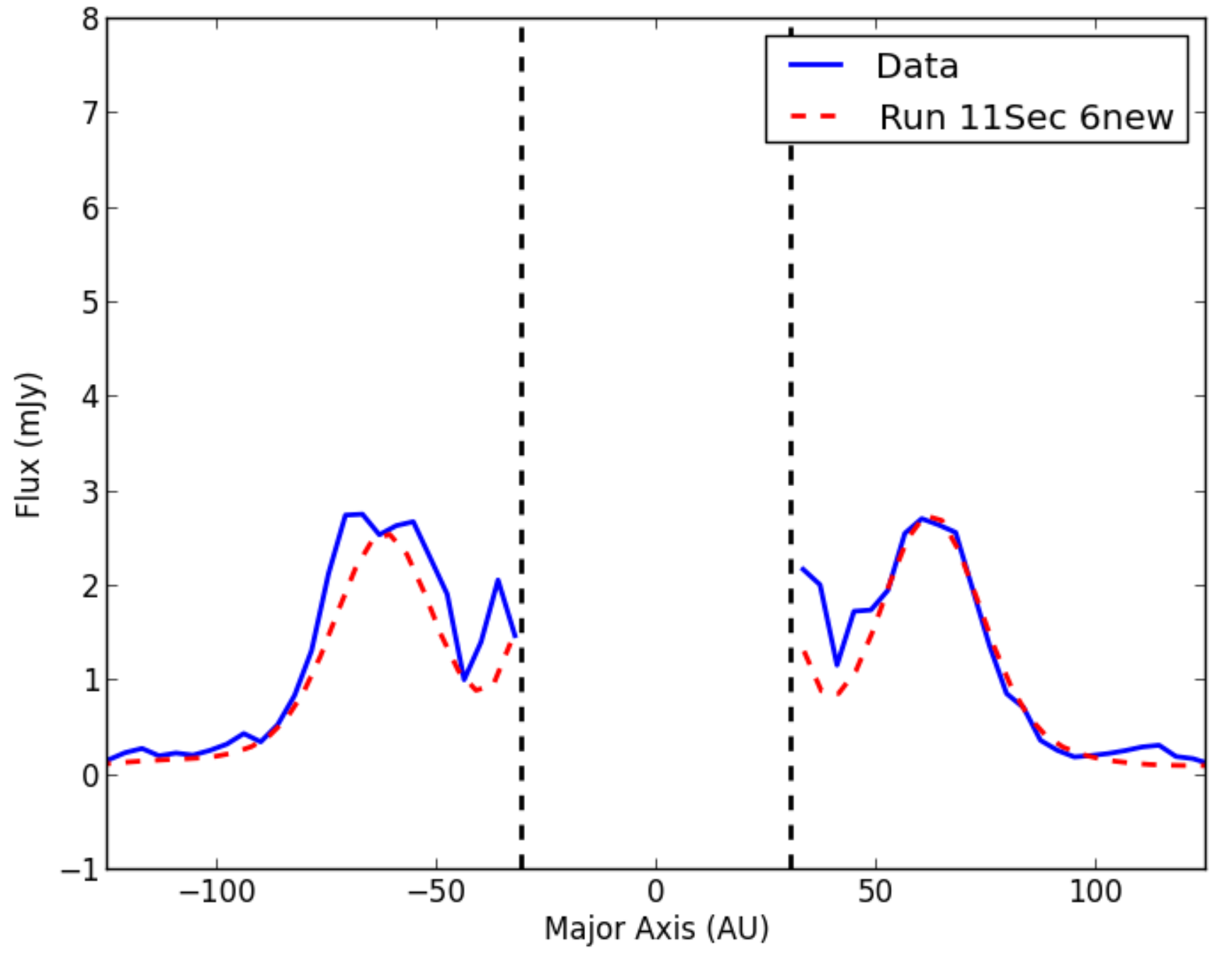}
\caption{Major axis crosscut of our 2011 H-band imagery data (PI image) compared to the best fit model (red-dashed line). The vertical dashed lines represent the inner working angle of $0.28"$}
\label{fig:model_major}
\end{figure}

\begin{figure}
\centering
\includegraphics[width=\columnwidth]{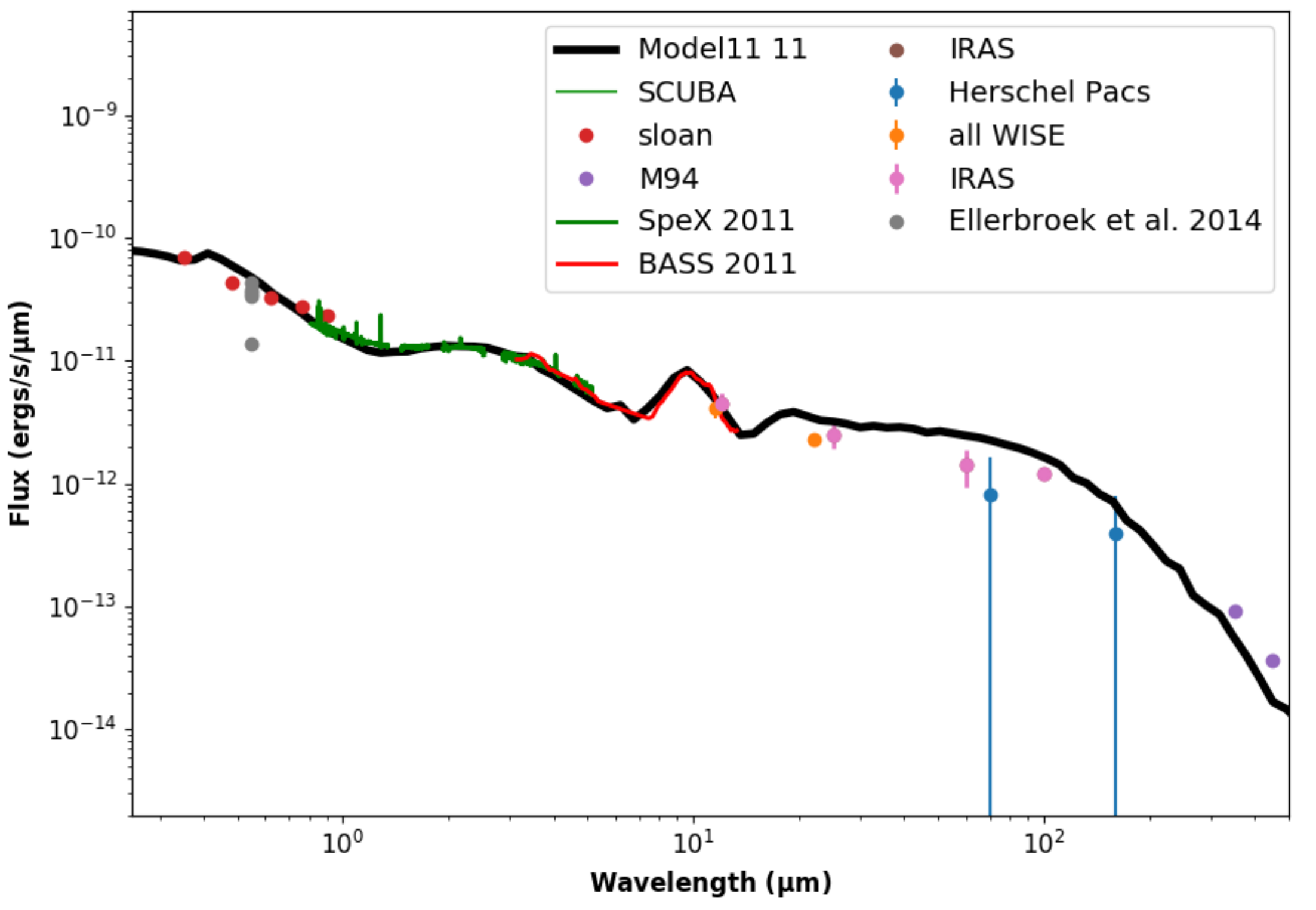}
\caption{The observed SED of HD 163296 is shown along with our best fit model SED (black line). The SpeX 2011 (red line) and BASS 2011 (teal line) data are from this work, as described in Section \ref{sec:obs}. The blue circles represent data from the AllWISE catalog \citep{wright2010}, the green circles are from the 2MASS All Sky Survey \citep{cutri2003}, and the purple circles are from IRAS point source catalog \citep{helou1988}. The gray circles depict V-band photometry and represent the historical minimum, 1-$\sigma$ below median flux, median flux, and 1-$\sigma$ above the median flux as reported by \citet{ellerbroek2014}. }
\label{fig:model_SED}
\end{figure}

Table \ref{tbl:model} lists the main parameters utilized in our best fit model, and Figure \ref{fig:trho} details the temperature and density profile of the disk in this model.  Figures \ref{fig:model_SED} and \ref{fig:model_major} show the SED and radial surface brightness profile along the disk major axis of our best fit model as compared to our 
observations.  We remark that our best fit model parameters are generally similar to those previously reported in the literature.  For example, our disk mass of 0.05 M$_\odot$ (Table \ref{tbl:model}) is similar to that measured by \citet{qi2011} (0.089 M$_\odot$) and \citet{isella2007} (0.12 M$_\odot$).

\begin{figure}
\centering
\includegraphics[width=\columnwidth]{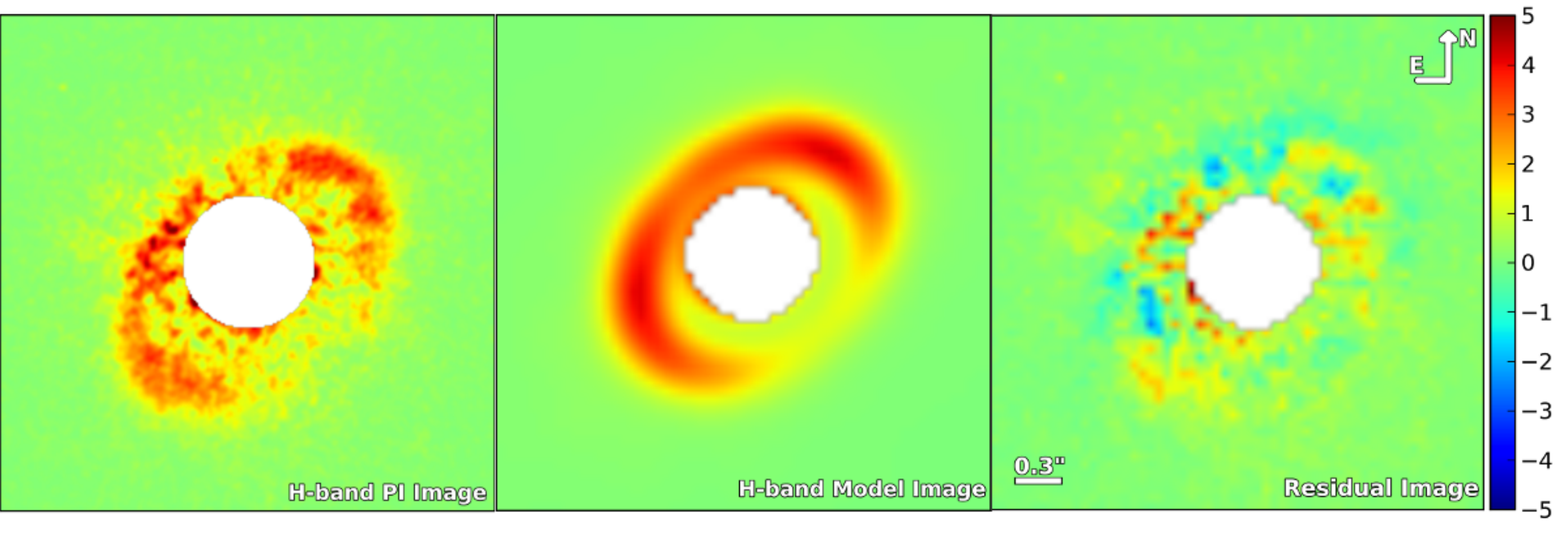}
\caption{Our 2011 H-band polarized scattered light image (left panel), the best fit model PI H-band scattered light image (middle panel), and the difference between the observed and model PI image (right panel) are shown. All three panels are displayed on the same linear scale, same spatial scale, and rotated such that North is up and East is left. The inner working angle is masked out with a white circle. Note that the PI image was binned to match the pixel scale of the model for the difference image.}
\label{fig:model_image}
\end{figure}

Our best fit model SED generally matches well with the contemporaneous spectroscopy and historical observations from optical to radio wavelengths (Figure \ref{fig:model_SED}).  Since the optical flux has been shown to be highly variable and we do not have contemporaneous optical photometry or spectroscopy, we do not know whether the modest model overestimation of the optical flux simply reflects that the star was at a high flux state in 2011. Additionally, our model reproduces the an on source extinction value of A$_V$ = 0.5 mag from \citet{ellerbroek2014} with a value of A$_V$ = 0.46 mag. We note that the observed versus model imagery comparison matches well along the NW side of the disk (right hand side of Figure \ref{fig:model_major}), while the model imagery is marginally too narrow along  the SE side of the disk (left hand side of Figure \ref{fig:model_major}).  This could be due to slight geometrical variations in the wall of the disk, causing the illumination of the SE-side of the 
disk to be broader.  We provide a full comparison of the observed H-band PI imagery and model imagery in Figure \ref{fig:model_image}.  Our model imagery reveals little scattered light beyond the bright ring and little to no scattered light within the gap of the disk, which matches the observed PI and Q$_\phi$ images.

\section{Analysis of SCExAO/CHARIS High-Contrast Near-Infrared Spectroscopy}\label{sec:charis}
\subsection{Methodology: Disk and Planet Forward-Modeling}
 Although none of the protoplanets/candidates reported from Keck/NIRC2 or ALMA are visible in our data, great care is needed to properly interpret  these non-detections and their implications.  For example, like HD 163296, HD 100546 has multiple imaged protoplanet candidates embedded in a bright, structured protoplanetary disk \citep{Quanz2013,Currie2015a}.   Follow-up claims of a spurious detection/non-detection of candidates around HD 100546 were faulty as shown in \citet[][]{Currie2017b}, in large part due to 1) incorrect spectrophotometric calibration and 2) a lack of forward-modeling of planet and disk signals.   
 
Contemporaneous near-IR spectra of HD 163296 allowed us to spectrophotometrically calibrate CHARIS data cubes (see Sect 2.1).  To properly understand our non-detections and derive upper limits at the candidates' locations, we then performed forward-modeling of our images, investigating the reduction of the total source signal and the biasing of its spatial intensity distribution due to processing.  This annealing results from self-subtraction of the source by itself and over-subtraction of the disk in ADI and SDI.  Our method follows that outlined in \citet{Currie2018b}, where we save the A-LOCI coefficients $\alpha$ and model the disk and planet signals as introducing a linear perturbation of value $\beta$, which provides an additional source of annealing \citep[see also][]{Brandt2013,Pueyo2016}.   We focus on the May 2018 data due to its higher quality.
 
 First, we explored the effect of disk on the non-detections of planetary companions, using forward-modeling to determine its annealing due to processing and its effect on any point sources located exterior, like the proposed companions from \citet{guidi2018} and \citet{teague2018}.   We started with the best-fit scattered light disk model described in Section \ref{sec:model}, which is drawn from our H-band scattered light imagery with Subaru/HiCIAO. We produced a total intensity (not scattered light) images in $J$, $H$, and $K$ passbands and interpolated the model images onto the CHARIS wavelength array and pixel scale. The model disk is slightly bluer than the combined light of the star+disk, with intrinsic colors of $J$-$H$, $H$-$K$ of $\sim$ 0.35 and $\sim$ 0.35. Note that the model was constructed based on a single passband (H-band), thus the model may not constrain the true color of the disk. The disk contrast with respect to the star on the forward-scattering side is typically $\Delta$M = M$_{disk/arcsec^{2}}$ - M$_{\star}$ $\approx$ 3.5--4.   
The visible trace of a disk may differ in total intensity vs. scattered light. Therefore, we slightly adjusted the model parameters to provide a better match to the forward-modeled disk image, specifically increasing the semimajor axis by 5\%.   

Second, we verified that an object consistent with the 6--7 $M_{\rm J}$ candidate from \citeauthor{guidi2018} would be detected in our data. We used standard hot-start evolutionary models from \citet{Baraffe2003}, adopting a planet age equal to the system age (5 Myr). This approach is intermediate between possible extremes that would yield higher and lower luminosities for a given planet mass.  While we assume a planet age of 5 $Myr$ when estimating mass limits, the age of a superjovian planet is likely much younger than that of the host star \citep{Currie2013}.   This is especially true for \textit{protoplanets}, which are nearing the end of their formation and thus much closer to $t$ $\lesssim$ 1 $Myr$ for any evolutionary model, where the planet luminosity is maximum.   The inferred limits adopting would be then substantially lower than those we report.   Conversely, we could adopt planet mass limits using the ``cold start" evolutionary models \citep[e.g.][]{Marley2007}.   However, recent literature casts serious doubt on the validity of the cold start model formalism, which relies on specific assumptions about the entropy of accreted material.   As shown by \citet{Berardo2017}, classic cold start conditions are extremely difficult to reach as the protoplanet will be substantially heated by the accretion shock, which will increase its entropy, resulting in hot start-like initial condition.  Furthermore, imaged planets for which we have derived dynamical masses -- $\beta$ Pic b, HR 8799 bcde \citep{Lagrange2010,Marois2010,Currie2011a,Snellen2018, Wang2018, Dupuy2019} -- are inconsistent with a cold-start evolutionary model. At the candidate's location in each data cube, we injected a planet whose temperature matches that expected for a 4 $M_{\rm J}$, 5 $Myr$ planet according to these models.
Although such a planet is predicted to be near the L/T dwarf transition ($T_{\rm eff}$ $\sim$ 1300 $K$), we assume a (cloudier) L dwarf spectrum drawn from the \citet{Bonnefoy2014} library, since annealing due to SDI will be stronger for such a spectrum.   Integrated over the CHARIS wavelength array, the broadband contrast of this planet with respect to HD 163296 is $\sim$ 8$\times$10$^{-6}$, about 2.5--3.5 times as high as the predicted contrast for the \citeauthor{guidi2018} companion using a cloudy planet atmosphere from \citet{Currie2011a}.

Finally, our forward-modeling calculation allowed us to compute radially-averaged, throughput-corrected broadband contrast curves.   As with our fake planet injection test, we used the \citet{Baraffe2003} models to map between planet mass and temperature.   To map between temperature and spectrum, we further used atmosphere models drawn from A. Burrows, adopting cloud prescriptions that provide reasonable fits to near-IR photometry for HR 8799 bcde and ROXs 42Bb, whose temperature, gravity, and masses cover most of our range \citep[][Currie et al. 2018 in prep.]{Currie2011a,Madhusudhan2011, Currie2014}.

\subsection{Results: Limits on Planets}

\begin{figure}
\centering
\includegraphics[width=0.49\columnwidth]{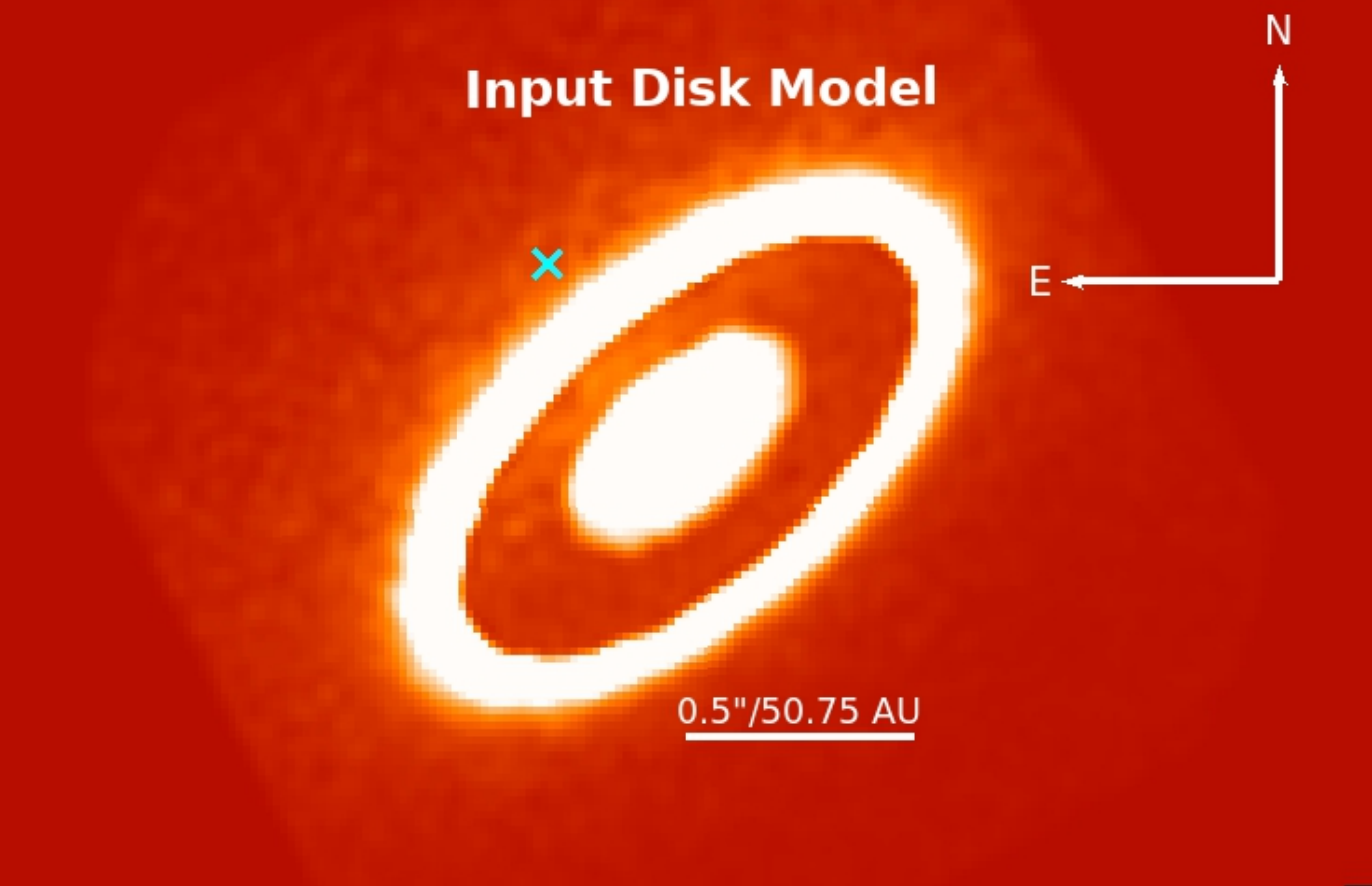}
\includegraphics[width=0.49\columnwidth]{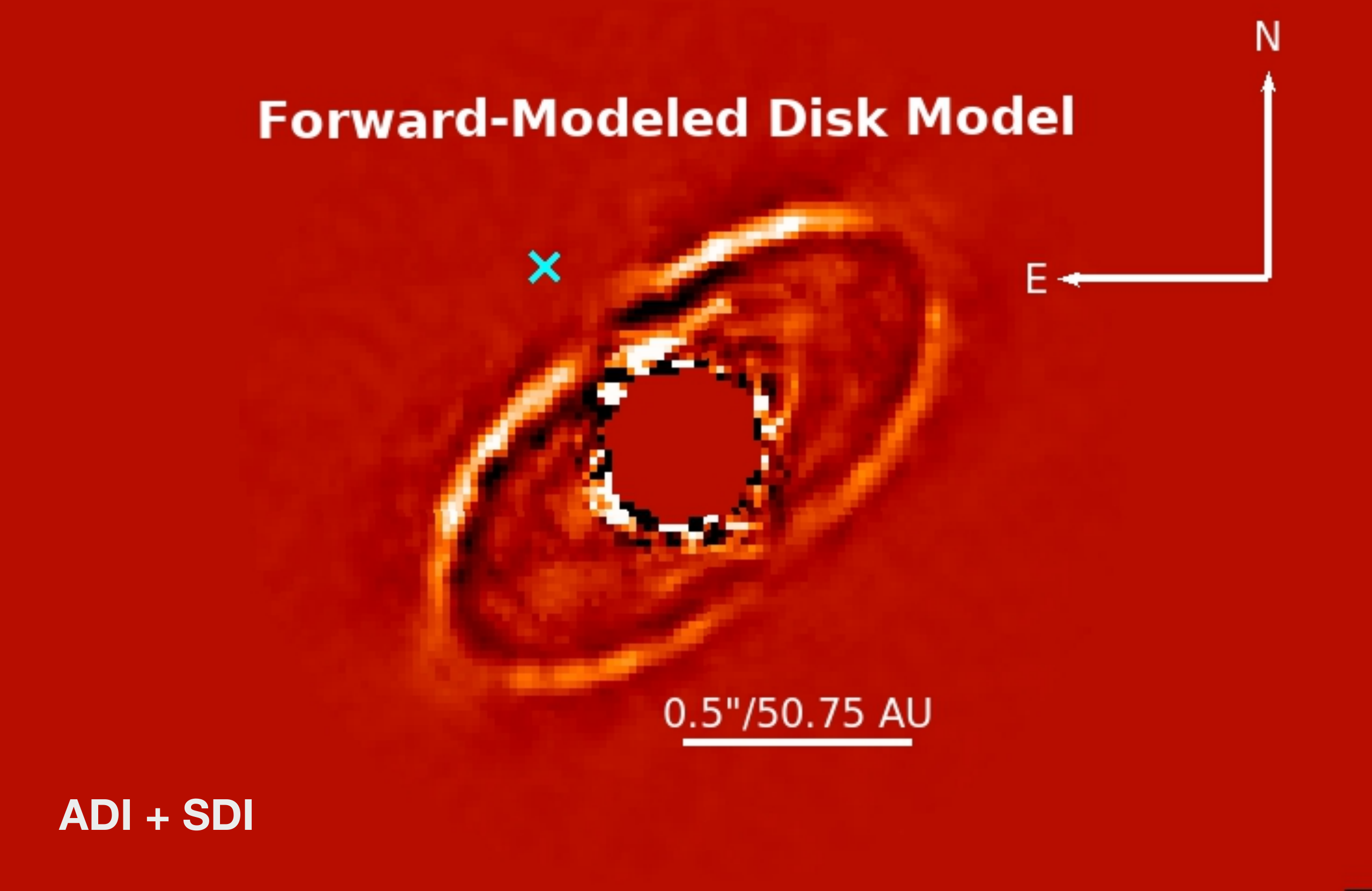}
\caption{(left) Broadband image of the best-fit synthetic disk model derived from polarimetry interpolated onto the CHARIS pixel scale and wavelength array and (right) forward-model of the disk after propagating its signal through ADI and SDI. The location of the proposed protoplanet candidate from \citet{guidi2018} lies well exterior to the azimuthal and radial self-subtraction footprints in the forward-modeled disk.   The images have been smoothed with a top-hat filter to more clearly reveal the trace of the disk: localized emission exterior to the disk is an artifact of this smoothing.}
\label{fig:fwdmoddisk}
\end{figure}

Figure \ref{fig:fwdmoddisk} shows the wavelength-collapsed image of the input disk (left panel) and output image after forward-modeling the disk through ADI and SDI (right panel).   While the disk in total intensity is more forward-scattering than the model based on polarimetry would predict and its brightness is $\sim$ 30\% higher, the model otherwise reproduces the CHARIS data and is sufficient for investigating the impact of self-subtraction on the forward-scattering side.  The proposed candidate from \citet{guidi2018} lies exterior to the main trace of the disk (cyan cross).   After processing, the candidate's location is free of negative self-subtraction footprints.   Inspection of the individual data cubes containing the disk model processed through ADI \& SDI likewise show a flat background.   At wider separations overlapping with the proposed candidate from \citet{teague2018}, the disk likewise leaves negligible residual effects.    

\begin{figure}
\centering
\includegraphics[width=0.49\columnwidth]{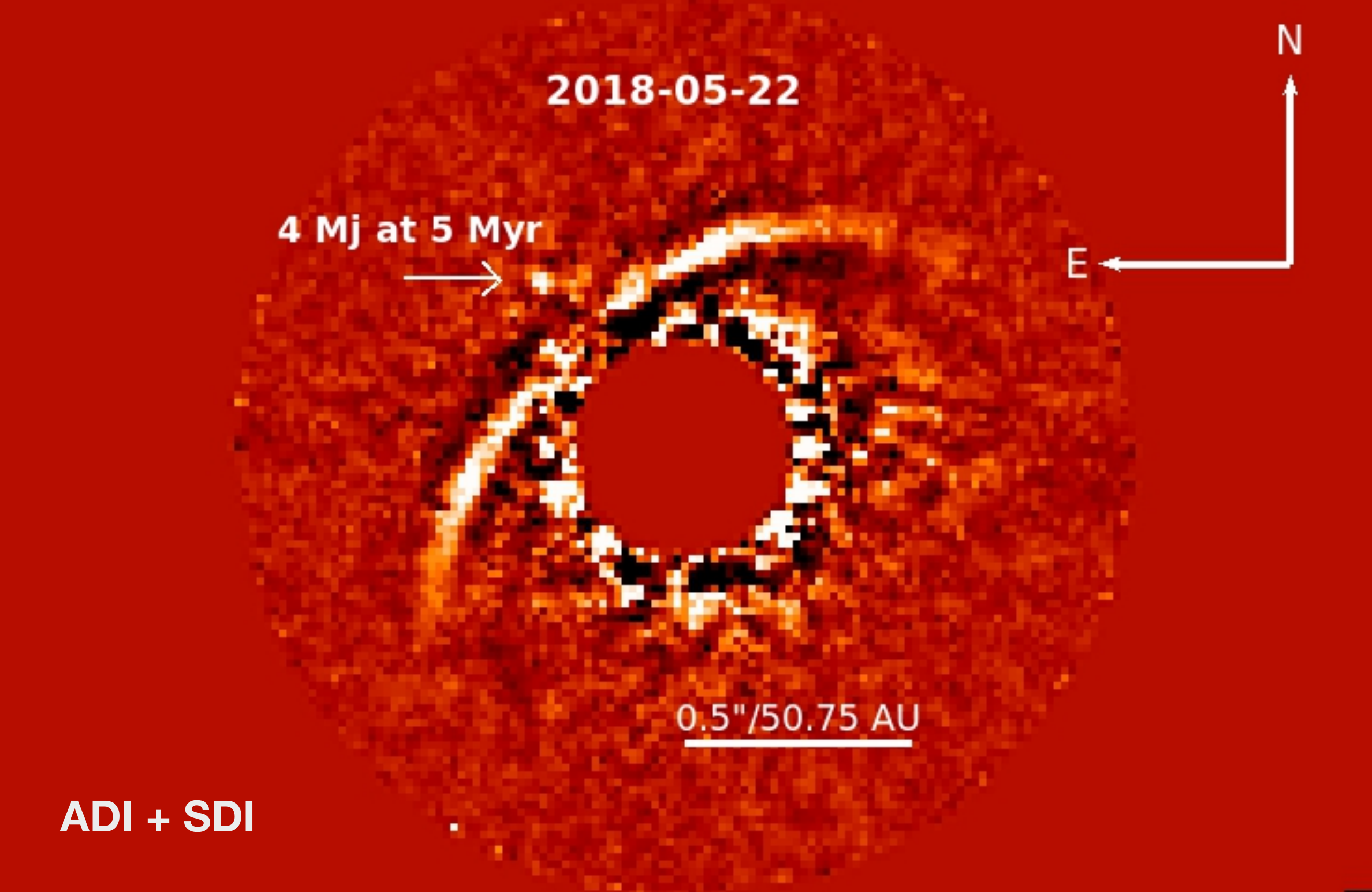}
\includegraphics[width=0.49\columnwidth]{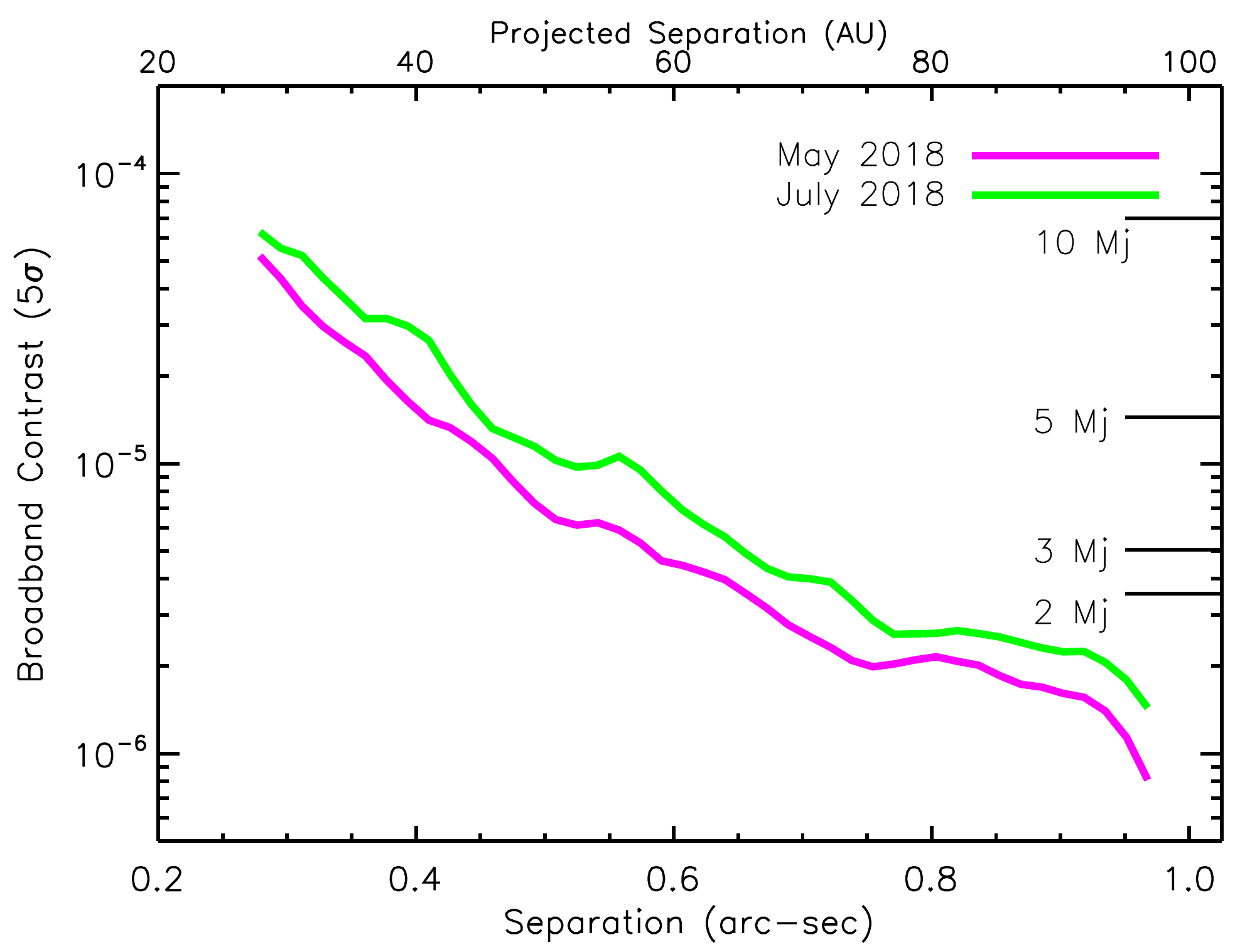}
\caption{(left) 2018 May broadband image with a 4 $M_{\rm J}$, 5 Myr-old planet injected into our observing sequence at the location of the candidate from \citet{guidi2018} ($\Delta$F $\sim$ 8$\times$10$^{-6}$) and propagating its signal through ADI and SDI.  Even with signal from the disk contributing to an estimate of the noise, the injected companion is detected at SNR $\sim$ 5. (right) Broadband contrast curve for the 2018 May and 2018 June data compared to broadband contrasts for 2--10 $M_{\rm J}$ planets assuming the Burrows atmosphere models.   The 5-$\sigma$ contrast at 0\farcs{}49 is in agreement with expectations based on our injected 4 M$_{\rm J}$ planet in the lefthand panel.   The contrast for a 1 $M_{\rm J}$ companion lies off the graph at $\Delta$F $\sim$ 3.7$\times$10$^{-7}$.}
\label{fig:planetlimits}
\end{figure}

As shown in Figure \ref{fig:planetlimits} (left panel), a 6--7 $M_{\rm J}$ candidate similar to the one proposed in \citeauthor{guidi2018} should have been detected in our data.   The fainter, even lower-mass (4 $M_{\rm J}$) candidate injected into our data is clearly visible.   While its SNR is formally $\sim$ 4.8, our inclusion of disk signal contributions leads our estimate of the noise to be conservative.    A planet corresponding to the \citeauthor{guidi2018} candidate ($\Delta$F $\sim$ 2.5 $\times$10$^{-5}$ would be even more decisively detected (SNR $\sim$ 15).  

Broadband contrast limits in the righthand panel of Figure \ref{fig:planetlimits} provide stringent limits on protoplanets covering the range probed with Keck/NIRC2 and ALMA.   At $\rho$ $\sim$ 0\farcs{}49, the azimuthally-averaged 5-$\sigma$ contrast limit is $\sim$ 8.5$\times$10$^{-6}$, in agreement with our expectations from the fake planet injection. If the \citeauthor{guidi2018} companion is real, it would then have to be redder than $H$ - $L_{\rm p}$ $\sim$ 3.5 to escape detection: redder than all directly-imaged planets except for the extreme L/T transition object HD 95086 b \citep{DeRosa2015}.   Over the separations just interior or close to the visible trace of the disk and comparable to the separation of the \citeauthor{guidi2018} companion -- $\rho$ $\sim$ 0\farcs{}4 (0\farcs{}7) along the minor (major) axis -- we can exclude planets with masses of 2--5 $M_{\rm J}$, assuming standard hot-start evolutionary models.   The CHARIS field encloses the possible location of the innermost companion proposed by \citet{teague2018}, which would lie at a projected separation of $r_{\rm proj}$ $\sim$ 83 au ($\rho$ $\sim$ 0\farcs{}82) along the major axis or $r_{\rm proj}$ $\sim$ 40 au ($\rho$ $\sim$ 0\farcs{}4) along the minor axis.   At these locations, our data rule out planets more massive than 5 $M_{\rm J}$ and $\sim$ 1.5 $M_{\rm J}$, respectively.   If located along the minor axis, the outermost proposed companion from \citet{teague2018} would be at $\rho$ $\sim$ 0\farcs{}65 with a mass less than $\sim$ 2 $M_{\rm J}$ according to our data. 

\section{Discussion}\label{sec:discussion}
 
\subsection{Previous optical-IR disk imaging}\label{sec:epoch}

HD 163296 has been observed numerous times across optical-IR bandpasses.  Here we briefly 
summarize some of the major results of those investigations, to compare and contrast with our new imagery.

Space-based optical imagery has been obtained in both white light (HST/STIS; \citealt{grady2000}) and broad-band filters (HST/ACS; \citealt{wisniewski2008}), tracing the 
disk out to 4$\farcs$4 (447 AU) and detecting HH knots.  Comparison of these data revealed evidence for significant variation ($\sim$1 magnitude) in the disk surface brightness, changes in the number of disk ansae visible over time, and changes in the relative brightness of features located in the NW and SE disk regions \citep{wisniewski2008}.  Unfortunately, none of these optical observations fully overlapped in wavelength coverage.  

Ground-based AO imagery of the system can be generally summarized into 3 categories.  First, a subset of observations clearly reveal the detection of the disk in scattered light, but the presence of residual AO speckle noise in the disk vicinity prevents a robust characterization of the surface brightness or detailed morphological structure of the disks (e.g. 2012 H-band imagery \citealt{garufi2014}; 2014 Ks-band imagery \citealt{garufi2017}).  Second, a subset of 
observations (e.g. 2012 Ks-band imagery; \citealt{garufi2014}) reveal the detection of an inclined ring structure extending out to 1$\farcs$03 (103 AU), where the intensity of scattered light is strongest along the major axis of the disk and is symmetrical about both sides of the disk major axis (NW-side and SE-side).  Third, a subset of observations 
(e.g. 2014 J-band \citealt{monnier2017}; 2016 H-band imagery \citealt{muro2018}) reveal clear evidence of this same inclined ring structure whose flux is both azimuthally asymmetric and not the strongest along the major axis.  In particular, the NW side of the major axis is brighter than the SE side of the disk in J-band GPI observations (see Figure 2, \citealt{monnier2017}), and the maximum flux from the disk is north of the major axis peaking on the NW side of the disk in these data.  The 2016 H-band VLT/SPHERE observations \citep{muro2018} also exhibit strong azimuthal asymmetry, with the NW-side of the disk along the major axis exhibiting 2.7x more scattered light than the SE-side of the disk along the major axis.  \citet{muro2018} used 3D radiative transfer modeling to suggest that this strong azimuthal asymmetry could be reproduced by including an inner disk component that was misaligned by 1$^\circ$ compared to the outer disk. 

\subsection{Evidence for time dependent azimuthal asymmetry} \label{sec:time}

Our 2011-epoch H-band imagery is consistent with the second category of disk appearance we discussed in Section \ref{sec:epoch}.  Namely, we observe a broken ring structure in H-band scattered light whose flux peaks along the major axis and exhibits clear symmetry between the NW- and SE-side of the disk.  Our 2011 epoch H-band data are thus clearly different than the 2016 epoch VLT/SPHERE H-band data, that show a 2.7x asymmetry between the NW- and SE-side of the disk \citep{muro2018}. 

To illustrate these differences, we scaled the peak flux along the major axis of the 2016 VLT/SPHERE data and present these data as dashed horizontal lines in our Figure \ref{fig:crosscuts}.  The 2.7x asymmetry about the major axis observed in the 2016 VLT/SPHERE data is clearly outside of the 3$\sigma$ errors of our 2011 data.  This obvious difference is also seen by comparing Figure 1 of \citet{muro2018} with Figure \ref{fig:images} in this paper.  We note that neither dataset exhibits evidence of large-scale gradients in their U$_\phi$ component, indicating that systematic artifacts are not the cause of this phenomenon.  We suggest that this is clear evidence that the system exhibits large changes in the appearance of its scattered light disk as seen in multi-epoch observations obtained with the same filter, and supports previous suggestions of this phenomenon as deduced from multi-epoch observations from similar, albeit not the same, filters \citep{wisniewski2008}. 

There are several mechanisms that could cause an azimuthal asymmetry of scattered light including an asymmetrical distribution of dust \citep{muro2018}, an inclined inner disk shadowing the outer disk \citep{muro2018}, a warped inner disk structure shadowing the outer disk \citep{sitko2008}, or dust ejected above the mid-plane of the disk that shadows the outer disk \citep{ellerbroek2014}. 

\citet{muro2018} suggested that an asymmetric distribution of dust in the system was unlikely, as no asymmetry was observed with ALMA \citep{isella2016}. \citet{muro2018} was able to replicate the azimuthal asymmetry they observed in their scattered light imagery by inclining the inner disk by 1$^\circ$ compared to the outer disk , which is consistent with previous near-IR interferometric observations \citep{tan2008,lazareff2017,setterholm2018}.  However, our 2011 epoch data reveal the presence of no azimuthal asymmetry along the major axis in the same filter bandpass as the 2016 SPHERE observations. An inclined inner disk is unlikely to precess significantly over a 5 year time-frame; hence, an inclined inner disk alone is unlikely to produce the observed significant azimuthal variations in the scattered light disk. Moreover, we have shown that we can reproduce the basic properties of both our contemporaneously obtained near-IR SED and H-band imagery with a model that does not include an inner inclined disk.  Thus, while the system could plausibly host an inclined disk, we suggest that this feature is unlikely to be responsible for producing the time-dependent azimuthal variations in the outer scattered light disk of the system.

We consider several other mechanisms that could explain the change in disk surface brightness seen in the system.  First, a warped inner disk structure, such as a puffed up inner disk wall \citep{turner2014}, could be shadowing the outer disk \citet{sitko2008}. If this disk warp were to dissipate or rotate azimuthally within a 2-3 year timescale, this could cause a change in illumination of the outer disk similar to that observed between the 2011 and 2016 epoch H-band datasets.  Dynamical simulations are needed to determine whether a substantial change in the appearance of a warped disk could occur on this short of a time-scale and lead to the amplitude of variable disk illumination observed.

\begin{figure}
\centering
\includegraphics[width=\columnwidth]{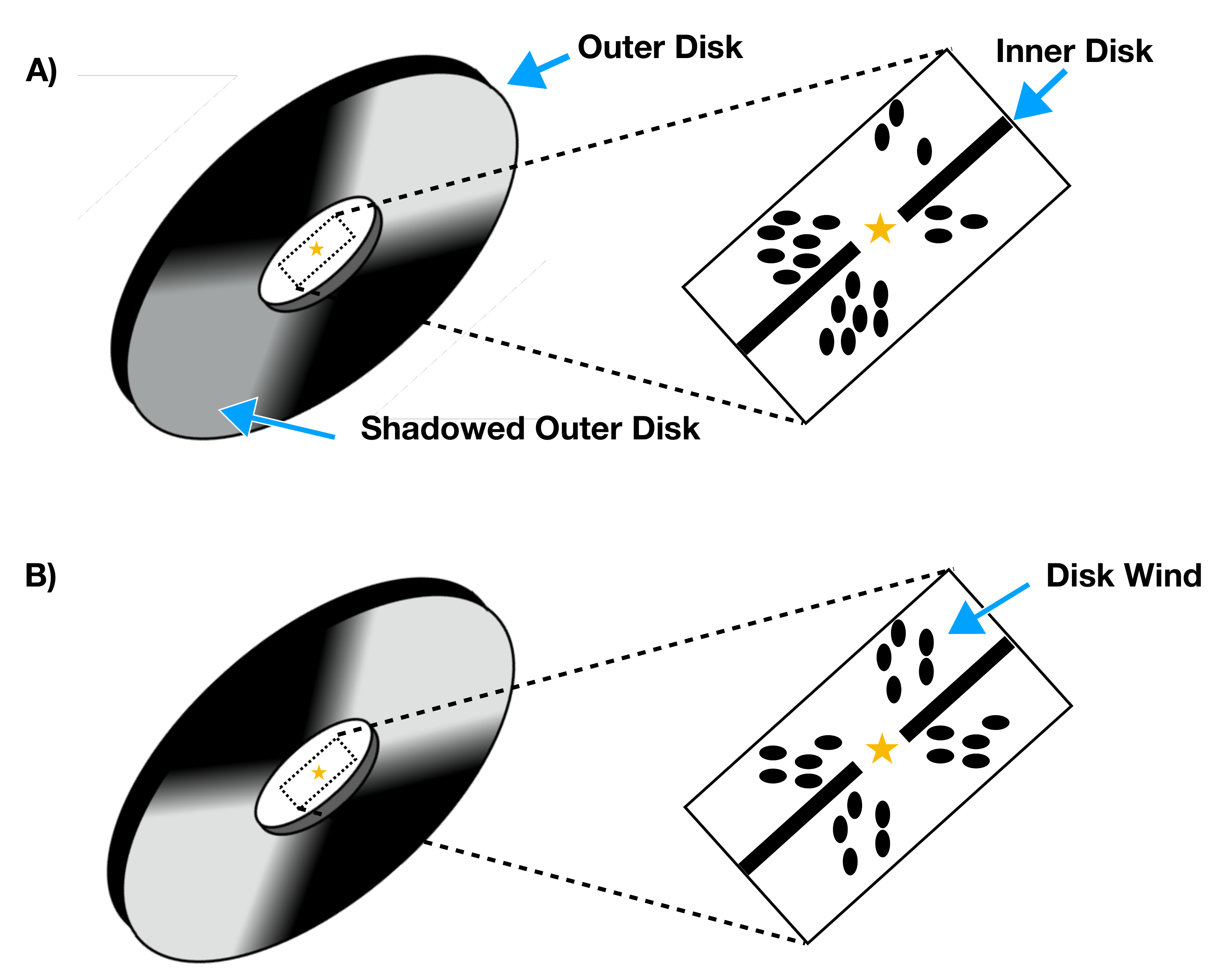}
\caption{Diagram of the disk wind model. A) shows the disk wind which is asymmetric which shadows the SE portion of the disk. B) shows a symmetric disk wind where the both sides of the disk are equally illuminated. The left hand side of the diagram shows the outer portion of the disk where the right hand side of the diagram shows a zoomed in version of the disk. The outer disk as been rotated and inclined to match the observed orientation of HD 163296 shown in Figure \ref{fig:images}.}
\label{fig:diagram}
\end{figure}

Second, this phenomenon could be caused by dust ejected above the mid-plane of the disk, which partially shadows the outer disk, as proposed by \citet{ellerbroek2014}. These dust ``clouds'' could differentially obstruct the illumination of the outer disk while they are between the star and the outer disk, as shown in Figure \ref{fig:diagram}.
We do have IR spectra that were obtained at a similar epoch to both our 2011 HICIAO data and the 2016 SPHERE data. The contemporaneous IR spectra cannot constrain the possible asymmetric nature of the dust clouds, but can constrain the total amount of dust in the disk wind when compared to our MCRT models. As shown in Figure \ref{fig:spex}, while both have the same flux around 0.9 $\mu$m, the 2011 epoch IR spectrum is brighter ($\sim$0.5 mag at K') around 2 $\mu$m than the 2016 epoch IR spectrum.  We remark that we can best reproduce the 2016 SED in our model by adopting a $\sim$2x lower envelope density, e.g. $9.0 \times 10^{-18} \frac{g}{cm^3}$, which corresponds to a lower circumstellar extinction in 2016 of A$_V$ = 0.1 mag. We predict that the 2016 epoch should be 0.4 mag brighter in the V-band compared to the 2011 epoch data, similar to the optical variability found by \citet{ellerbroek2014}. Since the interstellar extinction is consistent with AV = 0 mag, the observed reddening most likely originates from the system. Thus an asymmetric disk wind launching dust clouds can explains both the variable illumination of the outer disk (Figure \ref{fig:diagram}) and the reddening optical variability observed by our disk wind models and \citet{ellerbroek2014}.

If the system does have an inclined inner disk as suggested by \citet{muro2018} that during some epochs produces non-axisymmetric illumination of the outer disk (e.g. NW-side brighter than SE-side; 1998 HST/STIS \citealt{grady2000}, 2014 J-band \citealt{monnier2017}; 2016 H-band\citealt{muro2018}), the spatial distribution of any dust clouds elevated by a disk wind must also be non-axisymmetric to produce the observed epochs of axi-symmetric illumination of the outer disk (e.g. as seen in 2012 Ks-band imagery, \citealt{garufi2014}; 2011 H-band, this study) and the sole-epoch of observed non-axisymmetric illumination with the SE-side of the disk brighter than the NW-side (2004 HST/ACS \citealt{wisniewski2008}).  Future observations that simultaneously observe quiescent and wind events with contemporaneous optical and IR photometry and coronagraphic imagery could help to test whether shadowing by dust clouds could explain the observed behavior of the inner and outer disk of the system, and better parameterize the azimuthal distribution of such dust clouds.

\subsection{Model}

We were able to reproduce the basic properties of our contemporaneous near-IR spectra and scattered light H-band imaging with a 3D MCRT disk model, which approximated the features of a disk wind via an envelope. As seen in Figure \ref{fig:model_SED}, our model SED is consistent with the highest observed V-band flux that was reported by \citet{ellerbroek2014}, but we caution that the robustness of this agreement is uncertain as we do not have contemporaneous optical photometry. 

\citet{muro2018} also performed MCRT modeling of HD 163296, and compared their models to the ALMA dust continuum image from \citet{isella2016}, their own VLT/SPHERE image, and historical photometry and spectroscopy. They modeled all three gaps that were observed in the ALMA continuum image and introduced an inclined disk to explain the asymmetric scattered light flux observed with the VLT/SPHERE image as noted above \ref{sec:time}. Their model images and SED are well matched to their observed images and historical photometry and spectroscopy.  While they do not employ a disk wind model as we did (Section \ref{sec:model}, see Figure \ref{fig:diagram}), their model does not have a clear mechanism to explain the time dependent azimuthal asymmetries seen in near-IR scattered light images (Section \ref{sec:time}) or the optical-IR photometric and spectroscopic variability that has been observed \citep{sitko2008,ellerbroek2014}.  We caution that the inability of an inclined disk by itself to explain the observed time dependent azimuthal asymmetries observed in scattered light does not exclude the possibility that the system does in fact have an inclined disk. Due to limitations with Hochunk3D, we leave applying our disk wind model to the archival images and SEDs to future work.

\subsection{Scattered light features along the minor axis} \label{sec:Lprime}

We note that a deficit of scattered light is seen in the near-side of our disk imagery at a PA of 30$^\circ$, in both our binned imagery (Figure \ref{fig:theta_flux}) and our unbinned PI and Q$_{rot}$ images.  We caution that while this feature could be real, it is not uncommon to observe depolarization along the minor axis due to the residual presence of an un-corrected polarized halo.  Interestingly, this feature coincides with the disk brightness increase observed with the Keck/NIRC2 L'-band vortex coronagraph by \citet{guidi2018} and is located at a similar position angle, albeit closer to the host star, as the purported candidate planetary mass object reported by \citet{guidi2018}.  As noted by \citet{guidi2018}, this disk feature is located where forward scattering should be significant.  If the feature we observe at the similar disk position is astrophysical, the decreased amplitude of the feature in polarized intensity suggests that it could be polarized less than its neighboring disk material.  

\subsection{Limits on Protoplanets Orbiting HD 163296}
Our data improve the detection limits for protoplanets in thermal emission around HD 163296 compared to Keck/NIRC2 data from \citet{guidi2018}: from 5--7 $M_{\rm J}$ to now 2--5 $M_{\rm J}$, assuming standard hot-start evolutionary models, near the projected trace of the disk.  At wider separations covering the possible locations of the inner proposed candidate from \citet{teague2018} ($r_{\rm proj}$ $\sim$ 83 au/$\rho$ $\sim$ 0\farcs{}82), the limits have now improved from 4.5 $M_{\rm J}$ to 1.5 $M_{\rm J}$, the latter which is just slightly higher than the predicted mass of the companion (1 $M_{\rm J}$).  Limits for the outer \citeauthor{teague2018} candidate along the minor axis are likewise just slightly higher than the predicted mass (a limit of 2 $M_{\rm J}$ vs. a predicted 1.3 $M_{\rm J}$). Thus, at least for now, the ALMA-predicted protoplanet candidates are consistent with direct imaging constraints.

Our data appear to rule out the proposed, marginally-significant candidate identified from thermal IR data in \citet{guidi2018}. Using standard assumptions for planet atmospheres, our forward-modeling demonstrates we could have detected an even fainter planet at the location of the proposed candidate. For an assumed age of 5 $Myr$ and hot-start evolutionary models, the candidate is predicted to be 6--7 $M_{\rm J}$, while our radially-averaged contrast limits are significantly lower ($\sim$ 4--5 M$_{\rm J}$)\footnote{Note that any new age estimates for HD 163296 drawn from its GAIA-revised distance do not change our results. Comparisons to some isochrones may imply an older age (e.g. 7.6 $\pm$ 1.1 $Myr$; \citealt{vioque2018}). However, others (e.g. the MIST and PARSEC) isochrones imply ages comparable to or just slightly greater than 5 Myr (T. Currie, unpublished). These differences do not change the fact that the proposed HD 163296 companion should have been detected in our data under standard assumptions for planet atmospheres.}.

The simplest explanation for our conflicting results is that the NIRC2 candidate is instead residual, partially-subtracted speckle noise or partially-subtracted disk emission left over from processing.   Figure 1 of \citet{guidi2018} shows multiple emission peaks with a similar or slightly smaller spatial scale as the candidate (e.g. at the 2, 6, 7, and 8 o'clock positions just exterior to the masked region).   An even brighter, seemingly point source-like peak at nearly the same position angle in these data appears to be an artificially-enhanced region of the disk, which could have been mistaken for a point in shallower and/or higher background data.     Convolving the image with a gaussian kernel may further accentuate the point source-like appearance of these features\footnote{The large spatial scale of the residuals may also be traced to the PSF subtraction method used, which leverages on the Karhunen-Lo\'eve Image Projection (KLIP) algorithm with few KL modes retained \citep{Soummer2012}.  Compared to standard implementations of A-LOCI, KLIP with few KL modes retained may yield larger spatial scale residuals (T. Currie, unpublished).   This is especially true for KLIP implementations performing PSF subtraction in full annuli as in \citeauthor{guidi2018} instead of smaller wedge-shaped annular regions, since the subtraction is less local, in addition to constructing a low-rank approximation of the data set's covariance matrix.}.  The position of the candidate also coincides with the minor axis of a second ring of emission detected with ALMA.   Forward-modeling as performed in \citet{Currie2015a} could better clarify whether the candidate's morphology is consistent with an annealed point source or residual disk emission.

Alternatively, the candidate could be extremely red/underluminous in the near-IR and thus difficult to detect. If embedded in the disk, it would be preferentially extincted in the near-IR compared to the thermal infrared, as has been proposed for HD 100546 b \citep[][]{Currie2015a,Quanz2015}. It could also retain an extremely dusty/cloudy atmosphere characteristic of some young exoplanets near the L/T transition \citep{Currie2011a,DeRosa2015}, making it appear ``underluminous'' in the near-infrared. Follow-up thermal infrared imaging at $L_{\rm p}$ or $M_{\rm p}$ could provide a more decisive probe of these possibilities.

\section{Conclusions}
We report H-band polarimetric imagery of the HD 163296 system along with contemporaneous infrared spectra observations and near-IR extreme AO imaging in total intensity.  We find: \begin{itemize}

\item{Our 2011 H-band polarimetric imagery resolve a broken ring structure surrounding HD 163296 that peaks at a distance along the major axis of 0$\farcs$65 (66 au) and extends out to 0$\farcs$98 (100 AU) along the major axis.  Our non-detection of the inner disk component is driven by our inner working angle (0$\farcs$3, 30.5 au), and does not conflict with the detection of this component by \citet{monnier2017}.}

\item{Our 2011-epoch H-band imagery exhibits clear axisymmetry, with the NW- and SE-side of the disk exhibiting similar intensities.  Our 2011 epoch H-band data are thus clearly different than the 2016 epoch H-band data from VLT/SPHERE reported by \citet{muro2018}, that exhibit a strong 2.7x asymmetry between the NW- and SE-side of the disk.  These results indicate the presence of time variable, non-azimuthally symmetric illumination of the outer disk.}

\item{We were able to reproduce the basic properties of our contemporaneous near-IR spectra and spatially resolved H-band polarimetric imagery of the HD 163296 disk with a 3D MCRT disk model that approximated the features of a disk wind via an envelope and did not specifically require an inclined inner disk component.  We suggest that, while the system could plausibly host an inclined disk as suggested by \citet{muro2018}, such a component is unlikely to be responsible for producing the observed time-dependent azimuthal variations in the outer scattered light disk of the system.  We speculate that a variable, non-axisymmetric distribution of dust clouds elevated by a disk wind could produce the diversity of morphological appearances of the outer disk now reported in the literature for this system.}

\item{While our 2018 epoch SCExAO/CHARIS observations easily recovers the disk, they fail to recover the candidate 6--7 $M_{\rm J}$ protoplanet identified from Keck/NIRC2 data \citep{guidi2018}.  The Keck/NIRC2 detection is likely a residual speckle or a partially-subtracted piece of the disk; alternatively, this object could be a heavily embedded or particularly red/cloudy object only identifiable in the thermal infrared.}

\item{Assuming hot-start evolutionary models and a system age of 5 Myr}, our SCExAO/CHARIS detection limits for protoplanets in thermal emission around HD 163296 near the projected trace of the disk are 2--5 $M_{\rm J}$.  At wider separations, covering the possible locations of the inner proposed candidate from \citet{teague2018} ($r_{\rm proj}$ $\sim$ 83 au/$\rho$ $\sim$ 0\farcs{}82), our data lower the mass limit for detections from 4.5 $M_{\rm J}$ to 1.5 $M_{\rm J}$, which is still slightly higher than the predicted mass of the companion (1 $M_{\rm J}$).  Limits for the outer \citeauthor{teague2018} candidate along the minor axis are likewise just slightly higher than the predicted mass (a limit of 2 $M_{\rm J}$ vs. a predicted 1.3 $M_{\rm J}$).   The ALMA-predicted protoplanet candidates are currently still consistent with direct imaging constraints.
\end{itemize}

\acknowledgements
We acknowledge support from the NASA XRP program via NNX-17AF88G.  The authors recognize and acknowledge the significant cultural role and reverence that the summit of Mauna Kea has always had within the indigenous Hawaiian community. We are most fortunate to have the opportunity to conduct observations from this mountain.



\begin{table*}
\small
\centering
\caption{Results of ellipse fitting to PI H-band image}
\begin{tabular}{lc}
\hline
{Parameter} & {PI image Value}  \\
\hline
Major Axis of Disk (AU) & 58.01 $\pm$ 0.09  \\
Minor Axis of Disk (AU) & 48.4 $\pm$ 0.3  \\
Minor Axis offset (") & -0.0432 $\pm$ 0.0016  \\
PA (deg) & 132.2 $\pm$ 0.3 \\
Inclination ($^{\circ}$) & 41.4 $\pm$ 0.3 \\
\hline
\end{tabular}
\label{tbl:fit}
\end{table*}

\begin{table*}
\small
\centering
\caption{List of key best fit model parameters and estimates of the upper and lower bounds the parameter.}
\begin{tabular}{lccc}
\hline
{Parameter (Units)} & {Best fit Model} & {Lower Bound} & {Upper Bound} \\
\hline
Star Temperature (K) & 9250 & \nodata & \nodata \\
Star Radius (R$_\odot$) & 1.4 & 1.2 & 1.6 \\
Disk Mass (M$_\odot$)$^{(a)}$ & 0.05 & \nodata & \nodata \\
Fraction of Mass in Large Grain Disk & 0.9 & 0.8 & 0.95 \\
\hline
Inner Gap Radius (AU) & 29 & 20 & 32 \\
Outer Gap Radius (AU) & 59 & 55 & 62 \\
Large Grain Disk Minimum Radius (R$_{sub}$)$^{(b)}$ & 31.9 & 25 & 35 \\
Large Grain Disk Maximum Radius (AU) & 250.1 & \nodata & \nodata \\
Large Grain Disk Scale Height (R$_{sub}$)$^{(b)}$ & 0.11 & 0.08 & 0.13 \\
Large Grain Disk radial density exponent & 0.1 & 0.05 & 0.2 \\
Large Grain Disk scale height exponent & 0.16 &  & 0.18 \\
Small Grain Disk Minimum Radius (R$_{sub}$)$^{(b)}$ & 1.22 & 1.0 & 1.5 \\
Small Grain Disk Maximum Radius (AU) & 540.1 & \nodata & \nodata \\
Small Grain Disk Scale Height (R$_{sub}$)$^{(b)}$ & 0.11 & 0.08 & 0.13 \\
Small Grain Disk radial density exponent & 0.05 & & \\
Small Grain Disk scale height exponent & 1.25 &  &  \\
\hline
Envelope inner radius (R$_sub$)$^{(b)}$ & 0.41 & \nodata & \nodata \\
Envelope outer radius (AU) & 2.38 & \nodata & \nodata \\
Envelope Density ($\frac{g}{cm^3}$) & $4.0 \times 10^{-17}$ & $2.0 \times 10^{-17}$ & $6.0 \times 10^{-17}$ \\
Accretion (M$_\odot$) & $6.0 \times 10^{-7}$ & & \\
\hline
\end{tabular}
    \begin{tablenotes}
      \small
      \item (a) Disk mass value includes dust and gas. We assumed the gas to dust ratio is 100.
      \item (b) R$_{sub}$ is the sublimation radius with 1 R$_{sub}$ = 0.36 AU.  
    \end{tablenotes}
\label{tbl:model}
\end{table*}

\end{document}